\documentclass{tlp}

\jnlPage{1}{?}
\jnlDoiYr{2026}
\doival{10.1017/xxxxx}

\usepackage{hyperref} 
\usepackage{amssymb} 
\usepackage{stmaryrd} 

\newcommand{\codefont}{\tt\small}
\newcommand{\code}[1]{\mbox{\codefont{#1}}}
\newcommand{\ccode}[1]{``\code{#1}''}

\usepackage{listings}
\lstnewenvironment{curry}[1][]
  {\lstset{literate={->}{{$\rightarrow{}\!\!\!$}}3,#1}}{}
\lstnewenvironment{prolog}{}{}
\newcommand{\listline}{\vrule width0pt depth1.5ex}

\newtheorem{example}{Example} 
\newtheorem{theorem}{Theorem} 
\newtheorem{lemma}{Lemma} 

\renewcommand{\emptyset}{\varnothing}
\newcommand{\us}{\char95} 
\newcommand{\emptygoal}{\mbox{\tiny$\Box$}} 

\def\narr{\leadsto} 
\def\narrstar{\buildrel {*} \over \narr} 
\def\res{\vdash} 
\def\resstar{\vdash^{*}} 
\newcommand{\trans}[2]{\ensuremath{\left\llbracket #1
                       \right\rrbracket _{\cal #2}}}
\newcommand{\transt}[1]{\trans{#1}{T}} 
\newcommand{\translo}[1]{\trans{#1}{O}^{\respos}} 
\newcommand{\translc}[1]{\trans{#1}{C}^{\respos}} 
\newcommand{\transfr}[2]{\trans{#1}{F}^{#2}} 
\newcommand{\transf}[1]{\transfr{#1}{\respos}} 
\newcommand{\transd}[1]{\trans{#1}{D}^{\respos}} 
\newcommand{\respos}{{\cal R}}  
\newcommand{\numrespos}[1]{\#^\respos(#1)} 

\begin{document}

\title[Functional Logic Programming: Improving Logic Programs with Ease]
{Functional Logic Programming:\\ Improving Logic Programs with Ease\footnote{%
This is an extended version of a paper presented at the
34th International Symposium on Logic-based Program Synthesis and
Transformation (LOPSTR 2024),
invited as a rapid communication in TPLP.
The author acknowledges the assistance of the conference program chairs
Juliana Bowles and Harald S{\o}ndergaard.}}

\lefttitle{Michael Hanus}

\begin{authgrp}
\author{Michael Hanus}
\affiliation{Institut f\"ur Informatik, Kiel University, Germany. \\
        \email{mh@informatik.uni-kiel.de}}
\end{authgrp}

\maketitle

\begin{abstract}
Logic programming is based on defining relations.
Functions are often considered as syntactic sugar
which can be transformed into predicates so that
their logic is not used for computational purposes.
In this paper, we show that the use of functions
improves the operational behavior of logic programs
without loosing the flexibility of logic programming.
For this purpose, we present a systematic transformation from
predicates and goals into functions and nested expressions, respectively.
By evaluating functions in a demand-driven manner
and taking potential failures into account,
it is ensured that the execution of the transformed programs
will never require more steps than the original programs
but can decrease the number of steps---in the best case
reducing infinite search spaces to finite ones.
Thus, functional logic programming provides an adequate basis
to improve the operational behavior of logic programs
without changing their semantics.
This also shows that, from an operational point of view,
functional logic programming has clear advantages
compared to relation-based logic programming.

\medskip
\noindent
\textit{To appear in Theory and Practice of Logic Programming (TPLP)}

\end{abstract}


\section{Introduction}
\label{sec:intro}

Logic programming supports flexible programming techniques
by built-in non-determinism and free (logic) variables.
Predicates can be called with unknown arguments,
i.e., there are no fixed input and output positions,
in contrast to functional languages.
Since functions can be represented as predicates by adding
the result as a parameter, logic programming is often considered
as the more expressive programming paradigm \citep{Reddy84}.

Due to these considerations,
functions and nested function calls are often considered as nice syntactic
sugar which can be eliminated by translating them into predicates
and flattening nested expressions
\citep{BarbutiBelliaLeviMartelli84,CasasCabezaHermenegildo06,Naish91}.
A consequence of this traditional view is the fact
that one does not gain any operational advantage by
the presence of functions.
Since it is well known in functional programming
that demand-driven (lazy) evaluation supports new
programming techniques, as computing with infinite data structures
or modularity \citep{Hughes90},
there are also approaches to translate functions with
a demand-driven evaluation strategy by exploiting
coroutining in Prolog \citep{CasasCabezaHermenegildo06,Naish91}.
Since coroutining might influence completeness due to floundering,
these approaches have an ad-hoc flavor---they are useful for particular
examples, but general correctness results (soundness, completeness)
are not provided.

To improve this situation, one could also use an opposite approach.
Instead of transforming functions into predicates,
one could transform logic programs into programs
of a language with another operational semantics.
For instance, \citet{VanRoyHaridi23}
proposed to translate Prolog programs into Oz programs
where advanced features of Oz for concurrent and distributed
computations are used.
Although this is an interesting approach to move logic
programming techniques into the distributed world,
an operational improvement is not obtained by this transformation.
This is different in \citep{Hanus22TPLP} where
various transformations of logic programs into functional logic programs
are proposed.
Functional logic languages, such as Curry \citep{Hanus16Curry},
offer the same flexibility as logic languages
(soundness and completeness w.r.t.\ computing
with partial information).
In addition, optimality properties are known
for well-defined classes of programs (minimal number of computed
solutions, minimal number of evaluation steps \citep{AntoyEchahedHanus00JACM}).
The latter properties are due to the demand-driven evaluation
of functions, i.e., it is essential to keep nested functional expressions
instead of flattening them.

Due to these results, functional logic programming can be considered
superior compared to relation-based logic programming,
since any logic program can be transformed into a functional logic program
such that the operational behavior of the transformed program
is equal or better compared to the original logic program.
However, there is still one obstacle in this argument.
As shown in \citep{Hanus22TPLP}, this transformation
yields equivalent computations (identical answers, same number
of computation steps) if the functional logic program is
evaluated with an eager (strict) strategy.
When a lazy strategy is used, the transformed program
might be more efficient but there are also cases where
answers are computed which are not justified by the original logic program.
This could be the case if some subgoal fails in the logic program
but it is not evaluated (due to laziness) in the transformed program.
Thus, the transformation might obtain more efficient programs
but with logically different answers.

In this paper, we solve this problem by incorporating
the failure behavior into the transformation.
Due to this improvement, we obtain a method to
transform logic programs with the following properties.
\begin{itemize}
\item
The transformed programs always compute the same or more general
answers to a given goal.
\item
The evaluation of the transformed programs
is guaranteed equal or better than the evaluation of the
original programs: in the worst case, the same number
of steps are performed, but there are also cases where less steps
are required to compute the result.
\item
As a consequence, there are also cases where infinite
search spaces are reduced to finite ones.
\end{itemize}
Sloppily speaking, the improvements introduced by our transformation
are comparable to ``green'' cuts in logic programming.

To obtain this result, we extend the transformation
of \citep{Hanus22TPLP} by deciding the eager or lazy evaluation
of operations based on potential failures.
For this purpose, we carefully analyze the problem of
different answers between an eager or lazy evaluation
of transformed programs. Then we use an approximation
of the totality property of the transformed functions
to explicitly enforce strict evaluations at some points
in the generated programs.
If a subexpression is ensured to be totally defined,
it can be evaluated in a demand-driven manner,
otherwise its evaluation is enforced even if the result
is not immediately demanded.

We want to emphasize that the motivation of this work is
to show that functional logic programming has clear advantages
but no disadvantage compared to relation-based logic programming.
This is done by a systematic transformation which can be applied
to any logic program. The advantages of this transformation
are discussed and evaluated from an operational point of view.
Other aspects, like readability of programs, available extensions,
libraries, programming environments, are outside this work
since they are more difficult to grasp in a systematic manner.
For instance, Prolog offers various non-declarative features to influence
search, like the cut operator or predicates to examine classes of terms
(\code{var}, \code{nonvar}, \code{atom},\ldots) and restrict
the search for these, predicates with side effects, or libraries
containing specific algorithms for data structures and solving constraints.
As a consequence, most Prolog programs are not pure logic programs
although they use logic programming techniques in various ways.
It would be difficult to transform programs exploiting such features
into purely declarative programs.
On the other hand, contemporary functional logic languages,
such as Curry, offer declarative alternatives
to some of these features, like monadic input/output \citep{Wadler97},
set functions \citep{AntoyHanus09} to encapsulate
non-deterministic search, or default rules \citep{AntoyHanus17TPLP}.
Therefore, we argue that it is advantageous to define functions
instead of predicates
and use them with the flexibility as in logic programming.
As a consequence, one should write functional logic programs
directly from the beginning rather than writing logic programs
and transforming them into functional logic programs.
The work presented in this paper
justifies this argument by showing that nothing is lost
but there are essential improvements by this approach.

This paper is structured as follows.
After sketching the basics of logic and functional logic programming,
we review in Section~\ref{sec:lp2flp} existing methods to transform
logic into functional logic programs.
Section~\ref{sec:failing} discusses the incorporation of
failure information to obtain a transformation which
is correct w.r.t.\ the logical consequences of the logic program.
In order to improve the transformation,
we show in Section~\ref{sec:types} how the addition of types
can help to deduce more precise approximations of potentially
failing computations.
Section~\ref{sec:impl} sketches the implementation of our approach
which is evaluated in Section~\ref{sec:eval}.
Section~\ref{sec:related} discusses related work before we conclude.

This paper is an extended version of \citep{Hanus24LOPSTR}.
In addition to the original paper,
this paper contains a more detailed discussion of the approach,
a formal description of the transformation process,
more details about the implementation, and more examples
and benchmark data.

\section{Logic and Functional Logic Programming}
\label{sec:lp-flp}

In this section we briefly review some notions and features
of logic and functional logic programming.
More details can be found in \citep{Lloyd87}
and in surveys on functional logic programming
\citep{AntoyHanus10CACM,Hanus13}.

We use Prolog syntax to present logic programs.
\emph{Terms} in logic programs are constructed from variables ($X,Y,\ldots$),
numbers, atom constants ($c,d,\ldots$),
and functors or term constructors ($f,g,\ldots$) applied to
a sequence of terms, like $f(t_1,\ldots,t_n)$.
A \emph{literal} $p(t_1,\ldots,t_n)$ is a predicate $p$ applied
to a sequence of terms, and
a \emph{goal} $L_1,\ldots,L_k$ is a sequence of literals,
where $\emptygoal$ denotes the empty goal ($k = 0$).
\emph{Clauses} \code{$L$ :- $B$} define predicates,
where the \emph{head} $L$ is a literal and the \emph{body} $B$ is a goal.
A \emph{fact} is a clause with an empty body $\emptygoal$,
otherwise it is a \emph{rule}.
A \emph{logic program} is a sequence of clauses.

Logic programs are evaluated by SLD-resolution steps,
where we consider the leftmost selection rule here.
Thus, if $G = L_1,\ldots,L_k$ is a goal and
\code{$L$ :- $B$} is a variant of a program clause (with fresh variables)
such that there exists
a most general unifier\footnote{Substitutions, variants, and unifiers are
defined as usual \citep{Lloyd87}.}
(\emph{mgu}) $\sigma$ of $L_1$ and $L$,
then $G \res_\sigma \sigma(B, L_2,\ldots,L_k)$ is
a \emph{resolution} step.
We denote by $G_1 \resstar_\sigma G_m$ a sequence
$G_1 \res_{\sigma_1} G_2 \res_{\sigma_2} \ldots \res_{\sigma_{m-1}} G_m$
of resolution steps with $\sigma = \sigma_{m-1} \circ \dots \circ \sigma_1$.
A \emph{computed answer} for a goal $G$ is a substitution $\sigma$
(restricted to the variables occurring in $G$)
with $G \resstar_\sigma \emptygoal$.

\begin{example}\label{ex:plus-lp}
The following logic program defines a predicate \code{plus}
which relates two natural numbers in Peano representation, where \code{o}
represents zero and \code{s} represents the successor of a natural,
to its sum \citep{SterlingShapiro94}.
This predicate is used to define a predicate \code{plus3},
which relates three natural numbers to its sum,
and a predicate \code{thirdof} which is satisfied if the
second argument is a third of the first one.
\begin{prolog}
plus(o,Y,Y).
plus(s(X),Y,s(Z)) :- plus(X,Y,Z).$\listline$
plus3(X,Y,Z,R) :- plus(X,Y,XY), plus(XY,Z,R).$\listline$
thirdof(X,Y) :- plus3(Y,Y,Y,X).
\end{prolog}
For the goal \code{thirdof(o,T)},\label{ex:thirdof}
the answer $\{\code{T} \mapsto \code{o}\}$ is computed
but, after showing this answer,
Prolog does not terminate due to an infinite search space
(since it enumerates arbitrary large values for \code{Y}).
\end{example}
Functional logic programming languages \citep{AntoyHanus10CACM,Hanus13}
amalgamate the most important features
of functional and logic languages,
such as higher-order functions and lazy (demand-driven) evaluation
from functional programming
and non-deterministic search and
computing with partial information from logic programming.
The declarative multi-paradigm language Curry \citep{Hanus16Curry},
which we use in this paper,
is a functional logic language with advanced programming concepts.
Its syntax is close to Haskell \citep{PeytonJones03Haskell},
i.e., variables and names of defined operations
start with lowercase letters and
the names of data constructors start with an uppercase letter.
The application of an operation $f$
to $e$ is denoted by juxtaposition (``$f~e$'').

In addition to Haskell,
Curry allows \emph{free} (\emph{logic}) \emph{variables}
in program rules (equations) and initial expressions.
Function calls with free variables are evaluated by a possibly
non-deterministic instantiation of arguments.

\begin{example}\label{ex:plus-curry}
The following Curry program\footnote{%
The concrete syntax is simplified by omitting the declaration
of free variables, like \code{y} in the definition of
\code{thirdof}, which is required in Curry programs
to enable consistency checks by the compiler.}
defines the operations of Example~\ref{ex:plus-lp}
in a functional manner, where logic features
(like the free variable \code{y})
are exploited to define \code{thirdof}:
\begin{curry}
plus O     y = y
plus (S x) y = S (plus x y)$\listline$
plus3 x y z = plus (plus x y) z$\listline$
thirdof x | x =:= plus3 y y y
          = y
\end{curry}
Whereas \code{plus} and \code{plus3} are defined by equations
which are applied from left to right to evaluate expressions,
the function \code{thirdof} is defined by a conditional equation
where \ccode{|} introduces the condition.
A conditional equation can be applied to evaluate an expression
if the left-hand side matches the expression and the condition
can be evaluated to \code{True}.

The condition in the definition of \code{thirdof} contains
the infix operator \ccode{=:=} which denotes semantic unification:
an expression $e_1\;\code{=:=}\;e_2$ is evaluable to \code{True}
if $e_1$ and $e_2$ can be evaluated to
\emph{values} (i.e., expressions without defined functions)
$t_1$ and $t_2$, respectively,
and $t_1$ and $t_2$ are unifiable.
For instance, \code{plus\;(S\;O)\;x =:= S\;(S\;O)}
is solved by evaluating the left argument to \code{S\;x}
and binding variable \code{x} to \code{(S\;O)}.
In contrast to most Prolog systems, unification in Curry includes
the occur check so that \code{x\;=:=\;S\;x} is not unifiable.
This notion of equality is also called \emph{strict equality}
and used in functional logic languages with a non-strict (demand-driven)
semantics, see \citep{GiovannettiLeviMoisoPalamidessi91} for more details.
\end{example}
When a call to \code{thirdof} is evaluated, the functions \code{plus3}
and, therefore, also \code{plus} are called with arguments
containing free variables, like \code{y}.
In contrast to purely functional languages and similarly to
logic programming, this is allowed in functional logic languages.
Such calls are evaluated by instantiating free variables
in arguments so that a reduction step is enabled.
This process is called \emph{narrowing} \citep{Reddy85,Slagle74}
and defined as follows.
$t \narr_\sigma t'$ is a \emph{narrowing step}
if there is some non-variable position $p$ in $t$,
an equation $l \;\code{=}\; r$ (a variant of a program rule),
and an mgu $\sigma$ of $t|_p$ and $l$
such that $t' = \sigma(t[r]_p)$,\footnote{We use common notations
from term rewriting \citep{BaaderNipkow98}.}
i.e., $t'$ is obtained from $t$ by replacing
the subterm $t|_p$ by the equation's right-hand side and applying
the unifier.
Conditional equations $l \;\code{|}\; c \;\code{=}\; r$
are considered as syntactic sugar for the
unconditional equation $l \;\code{=}\; c \;\code{\&>}\; r$,
where the infix operator \ccode{\&>} is defined by
\code{True$\;$\&>$\;$x = x}.

A \emph{narrowing strategy} fixes the position of the subterm
selected in a narrowing step.
A survey on various narrowing strategies can be found in
\citep{Hanus94JLP}.
Curry is based on the
\emph{needed narrowing strategy} \citep{AntoyEchahedHanus00JACM}
which uses non-most-general unifiers in narrowing steps
to ensure the optimality of computations.
Needed narrowing is a demand-driven evaluation strategy,
i.e., it supports computations with infinite data structures \citep{Hughes90}
and can avoid superfluous computations so that it is
optimal w.r.t.\ the number of computed solutions and the length
of derivations (see \citep{AntoyEchahedHanus00JACM} for precise statements).
The latter optimality property is our motivation
to improve logic programs by transforming them into Curry programs.
This is not only of theoretical interest: the use of a
demand-driven narrowing strategy leads to a demand-driven
exploration of search trees so that this strategy
can reduce infinite search spaces occurring in logic programs
to finite ones in corresponding functional logic programs.
For instance, the evaluation of the expression \code{thirdof$\;$O}
has a finite computation space: the generation of larger numbers
for the first argument of \code{plus3} is avoided since there is
no demand for such numbers.

Curry has many more features which are useful for
application programming,
like \emph{set functions} \citep{AntoyHanus09} to encapsulate search,
and standard features from functional programming,
like modules and monadic I/O \citep{Wadler97}.
However, the kernel of Curry described so far should be sufficient
to understand the remaining contents.

\section{From Logic to Functional Logic Programs}
\label{sec:lp2flp}

Functional logic programming \citep{AntoyHanus10CACM}
amalgamates the features of functional and logic programming
in a single programming language.
Therefore, purely functional and purely logic programs
can be directly mapped into functional logic programs,
possibly by performing some syntactic adjustments.
For instance, there is a simple way
to transform logic programs into functional logic programs:
map each predicate into a Boolean function
and each clause into a (conditional) equation.
This \emph{conservative transformation} \citep{Hanus22TPLP}
does not change the operational semantics,
i.e., the length and structure of derivations,
since narrowing steps on Boolean functions
correspond to resolution steps.
Thus, there is no real advantage to perform such a simple transformation
so that we discuss more advanced transformations in this section.

\subsection{Functional transformation}
\label{sec:functrans}

In order to exploit the computational power of functional
logic languages, one can try to transform predicates
into non-Boolean functions by selecting some arguments
as results and generating function definitions according to
this selection.
\begin{example}\label{ex:plus-flp}
Consider the predicate \code{plus} defined in Example~\ref{ex:plus-lp}.
If the third argument is selected as a result argument
(as often intended in logic programs),
the clauses of \code{plus} can be
transformed into the following functional logic program:
\begin{curry}
plus O     y                  = y
plus (S x) y | z =:= plus x y = S z
\end{curry}
Thus, the result argument is moved to the right-hand side
of each defining equation and the bodies of clauses are
transformed into conditions where the unification operator \ccode{=:=}
is used to separate result arguments and the remaining call.
We call this the \emph{functional transformation} of logic programs.
Before we specify this transformation in detail,
we have to discuss a further aspect of this transformation.

Intuitively, the result argument is an argument of a given predicate
so that its value is determined by the values of the other arguments
in successful evaluations of this predicate.
Due to the fact that $n$-ary functions are often
represented as $(n+1)$-ary predicates in logic programs,
in many cases,
the result argument is the last argument of a predicate,
as in predicate \code{plus} defined in Example~\ref{ex:plus-lp}.
How can we find such an argument for arbitrary predicates?

Fortunately, in contemporary functional logic languages,
the strong intuition of result arguments, as sketched above, is not necessary.
In principle, any set of argument positions can be selected
as results.
For instance, if the first two arguments of \code{plus}
are selected as result arguments, the clauses of \code{plus}
can be transformed into the following program:
\begin{curry}
plus y                        = (O, y)
plus (S z) | (x,y) =:= plus z = (S x, y)
\end{curry}
Although this is not a function in a mathematical sense,
it is a valid definition in Curry:
it defines a \emph{non-deterministic operation}
which might deliver more than one result for a given argument.
For a given natural number $n$, it returns all splittings into two numbers
such that their sum is equal to $n$:
\begin{curry}
>  plus (S (S O))
(S (S O), O)
(S O, S O)
(O, S (S O))
\end{curry}
Non-deterministic operations, which can formally be
interpreted as mappings from values into sets of values \citep{GonzalezEtAl99},
are an important feature of contemporary functional logic languages.
The admissibility of non-deterministic operations
provides more freedom when transforming logic programs
into functional logic programs.
However, different selections of result arguments might lead to
different operational behaviors so that we discuss later
reasonable result argument selections.
\end{example}
The discussion above shows that the functional transformation
should be parametric w.r.t.\ positions of result arguments.
For this purpose, we assume that, for each $n$-ary predicate $p$,
there is an assignment
$\respos(p/n) \subseteq \{1,\ldots,n\}$
which defines the \emph{result argument positions}, e.g.,
$\respos(\code{plus}/3) = \{3\}$ or $\respos(\code{plus}/3) = \{1,2\}$.
In our implementation described in Section~\ref{sec:impl},
the definition of $\respos$ is automatically inferred
or it can be explicitly specified in the logic program
by specific directives.
For instance, the previous two result argument positions
can be specified by the directives
\begin{prolog}
:- function plus/3: 3.
\end{prolog}
and
\begin{prolog}
:- function plus/3: [1,2].
\end{prolog}
Hence, the result arguments are specified as a list of argument
positions where one can also write a single position
instead of a list containing a single element.

The first part of the definition of the functional transformation
specifies the transformation of terms.
Since terms in Prolog syntax
have a direct correspondence to data terms in Curry syntax,
the transformation of terms is just a change of syntax:
uppercase variables are mapped into lowercase,
lowercase constants and constructors are mapped into
their uppercase equivalents, and brackets in applications are moved outside.
This \emph{term transformation}, denoted by $\transt{\cdot}$,
maps a (Prolog) term into a Curry data term and is defined as follows:
\[
\begin{array}{@{}r@{~~}c@{~~}l@{~~~}l}
\transt{X} & = & x & \mbox{(variable)} \\
\transt{n} & = & n & \mbox{(number constant)} \\
\transt{c} & = & C & \mbox{(atom constant)} \\
\transt{f(t_1,\ldots,t_n)} & = & (F~\transt{t_1} ~\ldots~ \transt{t_n})
 & \mbox{(constructed term)}
\end{array}
\]
We denote the
\emph{functional transformation w.r.t.\ a result argument position mapping}
$\respos$ as a mapping $\transf{\cdot}$ from logic into
functional logic programs.
To define this transformation,
we use the following notation to split the arguments
of a literal into the result and the remaining arguments.
Consider a literal $l = p(t_1,\ldots,t_n)$ with the result argument positions
$\respos(p/n) = \{\pi_1,\ldots,\pi_u\}$,
where $\pi_i < \pi_{i+1}$ for all $1 \leq i \leq u-1$.
Then the \emph{tuple of result arguments} is defined as
\[
\translo{l} =
\left\{
\begin{array}{l@{\hspace{8.5ex}}l}
  \transt{t_{\pi_1}} & \mbox{if}~ u = 1 \\[0.5ex]
  \code{(}\transt{t_{\pi_1}},\ldots,\transt{t_{\pi_u}}\code{)} & \mbox{otherwise}
\end{array}
\right.
\]
If $\{\pi'_1,\ldots,\pi'_v\} = \{1,\ldots,n\} \setminus \respos(p/n)$,
where $\pi'_j < \pi'_{j+1}$ for all $1 \leq j \leq v-1$,
the \emph{call with the remaining arguments} is defined as
\[
\translc{l} = p\,\transt{t_{\pi'_1}} \,\ldots\, \transt{t_{\pi'_v}}
\]
Moreover, $\numrespos{l} = u$ denotes the number
of result arguments of the literal.

Based on these definitions,
we define the functional transformation $\transf{\cdot}$ as follows:
\[
\begin{array}{@{}r@{~}c@{~}l@{~~}l}

\transf{l} & = &
\left\{
\begin{array}{l@{\hspace{10.5ex}}l}
  \translc{l}
    & \mbox{if}~ \numrespos{l} = 0 \\[0.5ex]
  \translo{l} ~\code{=:=}~ \translc{l} & \mbox{otherwise}
\end{array}
\right.
 & \mbox{(literal)} \\[4ex]

\transf{l_1,\ldots,l_k} & = &
\left\{
\begin{array}{l@{\hspace{2ex}}l}
  \transf{l_1} ~\code{\&\&}~\ldots~\code{\&\&}~ \transf{l_k}
    & \mbox{if}~ k >  0 \\[0.5ex]
  \code{True} & \mbox{otherwise}
\end{array}
\right.
 & \mbox{(goal)} \\[4ex]

\transf{l.} & = &
\left\{
\begin{array}{l@{\hspace{13ex}}l}
  \translc{l} ~\code{=~True}
    & \mbox{if}~ \numrespos{l} = 0 \\[0.5ex]
  \translc{l} ~\code{=}~ \translo{l} & \mbox{otherwise}
\end{array}
\right.
 & \mbox{(fact)} \\[4ex]

\transf{l \code{~:-~} b.} & = &
\left\{
\begin{array}{l@{\hspace{4.2ex}}l}
  \translc{l} \code{~|~} \transf{b} ~\code{=}~ \code{True}
    & \mbox{if}~ \numrespos{l} = 0 \\[0.5ex]
  \translc{l} \code{~|~} \transf{b} ~\code{=}~ \translo{l} & \mbox{otherwise}
\end{array}
\right.
 & \mbox{(rule)} \\[4ex]

\transf{cls_1 \,\ldots\, cls_m} & = &
\transf{cls_1} \,\ldots\, \transf{cls_m}
 & \mbox{(program)}
\end{array}
\]
If a predicate $p$ with arity $n$ has no result arguments, i.e.,
$\respos(p/n) = \emptyset$, then it is transformed
into a Boolean function,
otherwise the result arguments are removed from the remaining arguments
and placed in the output position.

This formalizes the transformation of the predicate \code{plus}
defined in Example~\ref{ex:plus-lp} into the functional logic programs
shown in Example~\ref{ex:plus-flp}.
As a further example, predicate \code{plus3}
defined in Example~\ref{ex:plus-lp}
is transformed w.r.t.\ $\respos(\code{plus3}/4) = \{4\}$ into
\begin{curry}
plus3 x y z | xy =:= plus x y && r =:= plus xy z  = r
\end{curry}
The infix operator \ccode{\&\&} is the Boolean conjunction operator which
is predefined by
\begin{curry}
True  && x = x
False && _ = False
\end{curry}
Note that functions are not nested (this will be discussed below)
by the functional transformation $\transf{\cdot}$.
Due to the definition of \code{\&\&}, which requires
a Boolean value in the first argument so that it can be evaluated,
narrowing derivations w.r.t.\ the transformed program
have a strong one-to-one correspondence to resolution
derivations w.r.t.\ the original logic program
(apart from the single steps to evaluate each occurrence of \code{\&\&}).
This correspondence is independent of the selection of result arguments.
It is stated and proved in \citep{Hanus22TPLP} as follows.

\begin{theorem}[Correctness of the functional transformation]
\label{theo-correctness-functional}
Let $P$ be a logic program, $\respos$ a result argument position mapping
for all predicates in $P$, and $G$ a goal.
There is a resolution derivation $G \resstar_\sigma \emptygoal$
w.r.t.\ $P$
if and only if
there is a narrowing derivation $\transf{G} \narrstar_{\sigma} \code{True}$
w.r.t.\ $\transf{P}$.
\end{theorem}

This results means that there is no operational advantage of transforming
logic programs with the functional transformation
into functional logic programs.
Therefore, we present a refinement of this transformation.

\subsection{Demand functional transformation}
\label{sec:demand-functrans}

In order to get a real operational improvement of transforming
logic into functional logic programs,
one has to replace the strict equality operator
generated for literals occurring in goals
(see definition of $\transf{\cdot}$ for the case of literals)
by a local binding whenever possible.
Actually, this is possible
when the variables in the left-hand side of the unification operator
have no occurrences in result argument positions of other goal literals.
For instance, one can transform the rule
\begin{curry}
plus (S x) y | z =:= plus x y = S z
\end{curry}
into
\begin{curry}
plus (S x) y = S z   where z = plus x y
\end{curry}
Since variable \code{z} has only one occurrence (apart from its binding),
one can inline its definition, i.e., replace the binding of \code{z}
at this occurrence so that we obtain the typical functional definition
\begin{curry}
plus (S x) y = S (plus x y)
\end{curry}
As shown below, replacing strict equalities by local bindings
could yield less computation steps in a language with non-strict evaluation.

The extension of the functional transformation with possible
local bindings is called
\emph{demand functional transformation} \citep{Hanus22TPLP}
and denoted by $\transd{\cdot}$.
Its definition is identical to $\transf{\cdot}$
except for the translation of a literal in a goal.
Instead of a unification between the result arguments and the remaining call,
$\transd{\cdot}$ generates a local binding
\begin{curry}
$\translo{l}$ = $\translc{l}$
\end{curry}
(placed in the \code{where} part of the defining rule)
if $\numrespos{l} > 0$ and $\translo{l}$ is a tuple of
variables which do not occur in the rule's left-hand side
or in result argument positions of other goal literals.
The latter restriction avoids inconsistent bindings for a variable.

The operational behavior of this transformation depends
on the narrowing strategy of the functional logic language.
In a strict language, where arguments are evaluated before
jumping into the function's body (``call by value''),
there is no difference in the computed results
between the functional and the demand functional transformation,
as stated in the subsequent lemma.
Here, ``call-by-value narrowing'' is the narrowing strategy
where bindings are evaluated before the expressions in which
the bound variables occur. Thus, if such bindings are inlined,
this corresponds to strict or innermost narrowing.

\begin{lemma}[Correctness of the demand functional transformation w.r.t.\ call-by-value]
\label{lemma-correctness-demand-functional-cbv}
Let $P$ be a logic program, $\respos$ a result argument position mapping
for all predicates in $P$, and $G$ a goal.
There is a resolution derivation $G \resstar_\sigma \emptygoal$
w.r.t.\ $P$
if and only if
there is a call-by-calue narrowing derivation
$\transd{G} \narrstar_{\sigma} \code{True}$
w.r.t.\ $\transd{P}$.
\end{lemma}
\begin{proof}
This is a direct consequence of Theorem~\ref{theo-correctness-functional}.
The only difference between $\transd{\cdot}$ and $\transf{\cdot}$
is the translation of literals.
In a successful narrowing derivation w.r.t.\ $\transf{P}$,
a strict equality of the form $\translo{l}\;\code{=:=}\;\translc{l}$
is evaluated by evaluating its right-hand side $\translc{l}$
to some value (note that the left-hand side $\translo{l}$ is always
a value) and unifying this value with $\translo{l}$.
In a successful narrowing derivation w.r.t.\ $\transd{P}$,
this strict equality either directly occurs or it is present as
a local binding $\translo{l}\;\code{=}\;\translc{l}$.
In the latter case, a call-by-calue narrowing strategy
evaluates the bound expression $\translc{l}$ to some value
and binds $\translo{l}$ to this value (note that this binding is
always possible due to the restriction on variables in $\transd{\cdot}$),
i.e., the same instantiation
as in the narrowing derivation w.r.t.\ $\transf{P}$ is performed
so that both derivations perform the same computation steps
(at different positions).
\end{proof}
As a further example for the transformation $\transd{\cdot}$,
consider the $\transf{\cdot}$-transformed rule of \code{plus3} shown above.
The demand functional transformation replaces the strict equality operators
by nested local bindings so that we obtain
\begin{curry}
plus3 x y z = r   where xy = plus x y
                        r  = plus xy z
\end{curry}
If the bindings of variables \code{xy} and \code{r} are inlined, we obtain
the nested functional definition
\begin{curry}
plus3 x y z = plus (plus x y) z
\end{curry}
In principle, one can also inline bindings if more than one occurrence
of the bound variable exists, but this has the risk of duplicating
computations of the bound expression.
Inlining of local bindings is not essential since
Curry directly supports such bindings, similarly to functional languages,
but inlining can improve the readability of programs
so that we apply it in our examples.
Actually, this kind of inlining is the inverse of the normalisation
of functional programs presented by \citep{Launchbury93}
to specify a natural semantics for lazy evaluation.

As shown above, there is no operational difference
between programs transformed
by the functional and the demand functional transformation
if the transformed program is eagerly evaluated with a
call-by-value or innermost strategy.
This situation changes when the arguments are evaluated ``by need,''
as in Haskell or Curry
and discussed by \citet{HuetLevy91} and \citet{Hughes90}.
For instance, consider the predicate \code{signat}
defined by the clauses
\begin{prolog}
signat(o,o).
signat(s(X),s(o)).
\end{prolog}
Its demand functional transformation is\footnote{If it is not explicitly
mentioned, the last argument of a predicate is considered as
the result argument.}
\begin{curry}
signat O     = O
signat (S x) = S O
\end{curry}
Now consider the evaluation of the expression
\code{signat$\;$(plus$\;n_1\;n_2$)},
where $n_1$ is a big natural number.
An eager evaluation requires $n_1 + 2$ rewrite steps,
since the argument expression \code{(plus$\;n_1\;n_2$)}
is completely evaluated before evaluating \code{signat}.
In contrast, a non-strict language needs only two steps:
one step on \code{(plus$\;n_1\;n_2$)}, yielding the
expression \code{S\;(plus$\;\ldots$)}, followed by one application
of the second equation of \code{signat}.

Although it seems that the demand functional transformation
is the way to go, there is one potential problem
of this transformation: it might change the semantics, i.e.,
the set of computed solutions.
This could be the case if the evaluation of some subexpression
is not demanded and its evaluation would fail to yield a value.
This failure would be propagated in the original logic
program, but it might be ``hidden'' in the transformed program.
For instance, consider a decrement predicate
defined by the fact\label{ex:dec}
\begin{prolog}
dec(s(X),X).
\end{prolog}
and its application in the rule
\begin{prolog}
sig1(R) :- dec(o,X), plus(s(o),X,Y), signat(Y,R).
\end{prolog}
Due to the failure of the first subgoal, the goal \ccode{?- sig1(R).} fails.
However, the demand functional transformation yields (with inlining)
\begin{curry}
dec (S x) = x$\listline$
sig1 = signat (plus (S O) (dec O))
\end{curry}
so that a lazy evaluation of \code{sig1} yields the value \code{S$\;$O}.

One could argue that the latter result value is intended
due to the mathematical principle of ``replacing equals by equals.''
This point of view is taken in \citep{Hanus22TPLP}.
However, if we intend to use the transformation of logic programs
into functional logic programs in order to show that more efficient programs
can be obtained without changing the set of computed answers,
this difference is not acceptable.
Therefore, we develop in the next sections a slightly different
transformation which does not change the answer behavior
(except for computing more general answers)
but improves the operational behavior in many cases.

\section{Incorporating Failure Information}
\label{sec:failing}

We have seen in the previous section that the demand functional
transformation might yield a different answer behavior
if functions generated by the transformation are
partially defined, i.e., fail to compute a value
for particular argument values.
We call such functions \emph{failing}.
The reason could be a pattern matching failure as well as
an infinite loop.
Since failures due to infinite loops are seldom,
we later concentrate on the approximation of pattern matching failures.

If all functions occurring in an evaluation are totally defined,
i.e., always non-failing, then a demand-driven evaluation strategy,
like needed narrowing, computes only more general answers
compared to an eager strategy, as shown in the following result.

\begin{theorem}[Correctness of the demand functional transformation]
\label{theo-correctness-demand-functional}
Let $P$ be a logic program and $G$ be a goal.
Let $F = \transd{P}$ be the functional logic program and
$e = \transd{G}$ the expression
obtained by applying the demand functional transformation to $P$ and $G$,
respectively.
\end{theorem}
\begin{enumerate}
\item
If there is a resolution derivation $G \resstar_\sigma \emptygoal$ w.r.t.\ $P$,
then there is a needed narrowing derivation $e \narrstar_{\sigma'} \code{True}$
w.r.t.\ $F$ and a substitution $\varphi$ with $\sigma = \varphi \circ \sigma'$.
\item
If all functions in $F$ are totally defined and
there is a needed narrowing derivation $e \narrstar_{\sigma} \code{True}$
w.r.t.\ $F$,
then there is a resolution derivation $G \resstar_{\sigma'} \emptygoal$
w.r.t.\ $P$ and a substitution $\varphi$ with $\sigma' = \varphi \circ \sigma$.

\end{enumerate}
\begin{proof}
The proof is a direct consequence of the soundness and
completeness of needed narrowing \citep{Antoy97ALP,AntoyEchahedHanus00JACM}.

To prove 1., consider
a resolution derivation $G \resstar_\sigma \emptygoal$ w.r.t.\ $P$.
By Lemma~\ref{lemma-correctness-demand-functional-cbv},
there is a call-by-value narrowing derivation
$e \narrstar_{\sigma} \code{True}$ w.r.t.\ $F$ computing the same answer.
By soundness of narrowing,
$\sigma$ is a solution of the Boolean expression $e$ w.r.t.\ $F$.
By completeness of needed narrowing,
there is a needed narrowing derivation $e \narrstar_{\sigma'} \code{True}$
w.r.t.\ $F$ and a substitution $\varphi$ with $\sigma = \varphi \circ \sigma'$.

To prove 2., consider
a needed narrowing derivation $e \narrstar_{\sigma} \code{True}$ w.r.t.\ $F$.
Then, by soundness of needed narrowing,
there is a rewriting of $\sigma(e)$ to $\code{True}$.
This is not necessarily an innermost (call-by-value) rewriting, i.e.,
there are some function calls in this rewrite sequence
which might not be rewritten since their values are not needed.
In a call-by-value evaluation, all these function calls are evaluated.
Due to the assumption that all functions are totally defined,
they can always be evaluated to some value.
A call-by-value narrowing derivation might also bind some argument variables
which were not bound in the needed narrowing evaluation.
Thus, call-by-value narrowing might compute a more instantiated answer
$\sigma'$.
By Lemma~\ref{lemma-correctness-demand-functional-cbv},
there is a resolution derivation $G \resstar_{\sigma'} \emptygoal$ w.r.t.\ $P$.
\end{proof}

\begin{example}
Consider the expression \code{signat$\;$(plus$\;$x$\;$x)}
where \code{x} is a free variable.
Needed narrowing computes only the answer substitutions
$\{ \code{x} \mapsto \code{O}\}$ for the result \code{O} and
$\{ \code{x} \mapsto \code{S$\;$\us}\}$ for the result \code{S$\;$O},
whereas call-by-value narrowing computes
an infinite set of answer substitutions, i.e.,
$\{ \code{x} \mapsto n\}$ for every natural number $n$.
The same infinite set of answers is computed for the corresponding
Prolog program.
\end{example}
In order to safely use the demand functional transformation
w.r.t.\ the needed narrowing strategy,
one has to ensure that all functions
occurring in the evaluation steps are totally defined.
However, one has not to give up in the presence of
possibly failing operations: one has to enforce
the evaluation of such operations if they occur in the right-hand side
of a rule, since the corresponding predicates are also evaluated
in the logic program.
This can be obtained by the following modification
of the transformation.
If a rule's right-hand side contains an application $f~e$ and the evaluation of
the expression $e$ \emph{might fail} to compute a value,
then this application is replaced by
\begin{curry}
$f$ $\$$! $e$
\end{curry}
We call this modified transformation
\emph{fail-sensitive functional transformation}.
The predefined infix operator \ccode{$\$$!} denotes
a function application with a strict evaluation of the argument.
Thus, if the evaluation of $e$ fails, the evaluation of
\code{$f\;\$$!$\;e$} fails.

For instance, the fail-sensitive functional transformation
maps the predicate \code{sig1} defined in the previous section into
\begin{curry}
sig1 = signat $\$$! (plus (S O) $\$$! (dec O))
\end{curry}
so that the evaluation of \code{sig1} leads to a failure due to the
enforced evaluation of \code{(dec$\;$O)}.
Note that the usage of \ccode{$\$$!} is necessary at all places
where a potentially failing expression occurs and not only
where the failing expressions occurs first.
For instance, the slightly modified expression
\begin{curry}
signat (S (plus (S O) $\$$! (dec O)))
\end{curry}
evaluates to \code{S$\;$0} when evaluated by needed narrowing,
whereas
\begin{curry}
signat $\$$! (S $\$$! (plus (S O) $\$$! (dec O))
\end{curry}
fails, similarly to the corresponding logic program.

In order to perform the fail-sensitive functional transformation,
we need to know whether an operation is totally defined.
Obviously, this is undecidable in general so that
some approximation is required.
We can split this property into two parts:
termination and non-occurrence of failures.
Termination of rewrite systems or functional programs
is a well-studied topic so that various
techniques are available to approximate this property,
e.g., \citep{GieslEtAl11TOPLAS,LeeJonesBenAmram01}.
Actually, the Curry analysis framework CASS \citep{HanusSkrlac14}
also provides an analysis to approximate the termination of functions.
Therefore, we concentrate the subsequent discussion
to the problem to approximate fail-freeness.

An operation is called \emph{fail-free} \citep{Hanus18PPDP}
if its evaluation will never run
into an explicit failure (due to a pattern-match error).
A recent technique to approximate and verify fail-freeness
is a type-based approximation of the arguments describing values
for which a function call does not fail \citep{Hanus26SCICO}.
For this purpose, sets of concrete argument values
are approximated by some abstract type.
A simple but practically useful approximation are
depth-$k$ abstractions \citep{SatoTamaki84}.
For the case $k=1$, depth-1 types are described by subsets
of constructors representing all terms having one of these constructors
at the root. Moreover, the special value $\top$ describes
the set of all constructors occurring in the program
(a refinement based on type information will be discussed later).
An \emph{abstract call type} (or simply \emph{call type})
of an $n$-ary function $f$ is a sequence
$\alpha_1,\ldots,\alpha_n$ of abstract types such that any evaluation of
a call $f\,t_1 \ldots t_n$ does not fail if $t_i$ belongs to the
set represented by $\alpha_i$ ($i=1,\ldots,n$).
For instance, $\{\code{S}\}$ is a call type of the operation \code{dec}
($\emptyset$ would be another, less precise call type),
and $\{\code{O}, \code{S}\},\top$ is a call type of \code{plus}.
Since $\{\code{O}, \code{S}\}$ are the only constructors
in this program, we can also write this call type as $\top,\top$.
A \emph{trivial call type} is a sequence of $\top$s.

A method to infer call types is described in \citep{Hanus26SCICO}.
It is based on approximating the input/output behavior of operations
and using this information to infer call types by considering
the patterns and the operations occurring in the right-hand sides
of the rules. Since these operations might demand for refined
call types, the entire inference is a fixpoint computation
on the complete program.
For instance, the initial call type of
\begin{curry}
plusDec x y = plus x (dec y)
\end{curry}
is $\top,\top$ (since there are no restrictions in the left-hand side
due to patterns). The call \code{(dec$\;$y)} demands that the second
argument \code{y} has type $\{\code{S}\}$ so that the initial call type
is refined to $\top,\{\code{S}\}$.
Analyzing \code{plusDec} again with this refined call type,
\code{plusDec} can be proved as fail-free.

An operation can have a trivial call type even if it calls
potentially failing operations. For instance, consider the operation
\begin{curry}
decPlusS n = dec (plus (S n) n)
\end{curry}
The tool described in \citep{Hanus26SCICO} infers the trivial call type
for \code{decPlusS} since the first argument of \code{plus}
ensures that \code{dec} is always called with an \code{S}-rooted term.

The practical evaluation shows that only a few operations
have non-trivial call types in larger programs,
since most operations are defined by complete pattern matchings
on their input type.
However, this demands for the consideration of the intended types
of operations, as discussed next.

\section{Type-Based Translation}
\label{sec:types}

Although Curry and other contemporary functional logic languages
(e.g., TOY \citep{Lopez-FraguasSanchez-Hernandez99}) are strongly typed,
we have not considered type information in the
fail-sensitive functional transformation.
In principle, this is not necessary since types of functions
are automatically inferred.
Since data constructors need to be declared in Curry,
the implementation of our transformation
adds a single definition of a type \code{Term}
containing all constructors occurring in the program.

\begin{example}\label{ex:numnodes}
The transformation of the examples shown in the previous section,
where natural numbers in Peano representation are used as values,
adds the type declaration
\begin{curry}
data Term = O | S Term
\end{curry}
Consider that we extend the previous logic programs by a definition
of a predicate to relate a binary tree with the number of its nodes,
like
\begin{prolog}
numnodes(leaf(_),s(o)).
numnodes(node(M1,M2),s(N)) :-
        numnodes(M1,N1), numnodes(M2,N2), plus(N1,N2,N).
\end{prolog}
Then the transformation generates the data type declaration
\begin{curry}
data Term = O | S Term | Leaf Term | Node Term Term
\end{curry}
Since the logic program is not strongly typed,
it is not ensured that \code{plus} is called only with 
natural numbers as arguments.
Thus, the call type inferred for \code{plus} is no longer trivial.
This is correct since the call
\code{plus\;(Leaf\;O)\;O} is allowed but fails.
\end{example}
In order to make the failure analysis more precise,
we extend the transformation by the inclusion of type declarations.
Although there is long history about types in logic
programming \citep{MycroftOKeefe84,Pfenning92}
and also some interest to add types
to Prolog programs \citep{SchrijversSantosCostaWielemakerDemoen08},
there is no general agreement about its syntax and structure.
For instance, CIAO-Prolog \citep{HermenegildoEtAl12}
allows the definition of regular and Hindley-Milner types,
or \citet{BarbosaFloridoSantosCosta21} describe
a tool to infer types from logic programs.
In our transformation tool,
we support the explicit definition of polymorphic
algebraic data types in Prolog syntax, like
\begin{prolog}
:- type nat = o ; s(nat).$\listline$
:- type tree(A) = leaf(A) ; node(tree(A),tree(A)).
\end{prolog}
These will be translated into definitions in Curry syntax:
\begin{curry}
data Nat = O | S Nat$\listline$
data Tree a = Leaf a | Node (Tree a) (Tree a)
\end{curry}
With these definitions, the following types are inferred
for the transformed operations \code{plus} and \code{numnodes}:
\begin{curry}
plus :: Nat -> Nat -> Nat$\listline$
numnodes :: Tree a -> Nat
\end{curry}
Based on these types, both operations have trivial call types
so that they are fail-free.
This demonstrates that type information is useful to improve
the transformation of logic programs into functional logic programs.
In principle, it is not necessary.
In the worst case, missing type information might have the effect
that all operations are potentially failing so that the
translated program is eagerly evaluated without any advantage
compared to the original logic program.

\section{An Automatic Transformation Tool}
\label{sec:impl}

We presented in the previous sections a method to transform
logic programs into functional logic programs which ensures
that the transformed programs compute the same or more general answers
than the original logic programs.
Moreover, we showed that the operational behavior of the
transformed programs, i.e., the number of computation steps,
can be better or is, in the worst case, identical to the
original logic programs.
Although this shows the principle advantages of functional logic programming
compared to relation-based logic programming,
it is also interesting to see whether real improvements can be obtained
so that the worst case is not the regular case.
In order to evaluate our approach on arbitrary logic programs,
we implemented it in an automatic transformation tool.
This section describes the main aspects of the implementation
of this tool.

\subsection{Inferring result argument positions}

Our method to transform logic programs into functional logic programs
is parametric w.r.t.\ the arguments of predicates to be considered
as results.
As discussed in Section~\ref{sec:functrans},
any set of argument positions can be selected as results
to obtain a semantically correct transformation.
However, the concrete selection of result argument positions
can influence the operational behavior.
For instance, if the transformed program defines functions
in the mathematical sense, a demand-driven evaluation
of nested function calls might reduce the number of evaluation steps,
whereas a transformation into non-deterministic operations
might keep the number of evaluation steps.
However, there are also examples where non-deterministic operations
contribute to reduced search spaces.
For instance, the complexity of the classical permutation sort
can be improved by defining the computation of permutations
as a non-deterministic operation \citep{GonzalezEtAl99,Hanus13}.

Our transformation tool supports the explicit definition
of result argument position as shown in Section~\ref{sec:functrans}.
In order to support a fully automatic transformation
without requiring such explicit definitions,
our tool uses a heuristic if such definitions are not given.
The main idea of this heuristic comes from the theory
of term rewriting \citep{HuetLevy91} and
functional logic programming \citep{AntoyEchahedHanus00JACM}
where it is known that evaluations can be optimal for particular
classes of programs:
if function definitions are
\emph{inductively sequential} \citep{Antoy92ALP},
needed narrowing computes a minimal set of solutions and
successful derivations have minimal length \citep{AntoyEchahedHanus00JACM}.

A precise definition of ``inductively sequential''
is rather complex and can be found in \citep{Antoy92ALP,Hanus22TPLP}.
Therefore, we only sketch its idea.
Intuitively, a function definition is inductively sequential
if the left-hand sides of the rules of each function
contain arguments with a unique case distinction on different
data constructors.
Many predicate definitions of logic programs contain
a single argument which allows a unique case distinction
between all clauses,
e.g., the first argument in the predicates \code{plus}, \code{signat}
or \code{numnodes} shown above.
Due to this property, Warren's Abstract Machine \citep{Warren83}
has specific instructions to support efficient code generation for
indexing on the first argument \citep{Ait-Kaci91}.
In general, there are also predicates requiring other or
more than one argument for a unique selection of a matching rule.
For instance, \citet{SterlingShapiro94} define
a logic program for Ackermann's function as follows:\label{ex-ackermann}
\begin{prolog}
ackermann(o,N,s(N)).
ackermann(s(M),o,V) :- ackermann(M,s(o),V).
ackermann(s(M),s(N),V) :- ackermann(s(M),N,V1), ackermann(M,V1,V).
\end{prolog}
The first argument distinguishes between the cases of an atom \code{o}
and a structure \code{s(M)}, but, for the latter case,
two rules might be applicable. Hence, the second argument is necessary
to distinguish between these rules.
Therefore, $\{1,2\}$ is the set of inductively sequential argument positions
for \code{ackermann} whereas \code{plus}, \code{signat}
or \code{numnodes} have $\{1\}$ as inductively sequential argument positions.
In all cases, the last argument can be selected as the
result argument position.

Due to these considerations, our tool uses the following heuristic
to infer the result argument positions for a $n$-ary predicate $p$:
\begin{itemize}
\item[]
If a set $D$ of inductively sequential argument positions exists,
$R = \{1,\ldots,n\} \setminus D$ is not empty,
and $m$ is the maximum value of $R$,
then $\respos(p/n) = \{m\}$, otherwise $\respos(p/n) = \emptyset$.
\end{itemize}
Thus, a predicate is transformed into a function only if there are
some inductively sequential arguments and some other arguments.
In this case, we select the maximum argument position,
since often the final argument is interpreted as the result
in predicates representing functions.
On the other hand, the predicate
\begin{prolog}
some(true,X).
some(X,true).
\end{prolog}
has no inductively sequential arguments so that it is translated
into a Boolean function.

Note that one can always override this heuristic by adding
\code{function} directives as shown in Section~\ref{sec:functrans}.

\subsection{Implementation of the transformation tool}
\label{sec:toolimpl}

Our transformation tool is implemented as an extension of the tool
presented in \citep{Hanus22TPLP} where type and failure information
is included in the transformation process.\footnote{The tool, available at
{\scriptsize\url{https://cpm.curry-lang.org/pkgs/prolog2curry.html}},
is implemented as a Curry package for easy installation.
A script with all required tools is also
available as a docker image at
{\scriptsize\url{https://hub.docker.com/r/currylang/prolog2curry}}.
}
Types can be declared in source programs as
discussed in Section~\ref{sec:types}.
They are directly mapped into corresponding data definitions
in Curry.
The remaining constructors occurring in the logic program
but not mentioned in a type declaration are declared
in a single \code{Term} type, as shown in Example~\ref{ex:numnodes}.

Predicates are mapped into Curry functions without type annotations
so that type signatures are inferred by the Curry system.
The entire transformation process is performed in the following steps:
\begin{enumerate}
\item
The given logic program is translated into a Curry program
based on the demand functional transformation $\transd{\cdot}$
as described in Section~\ref{sec:demand-functrans}.
The result argument positions are either
explicitly declared in the given logic program
or inferred as described above.
\item
The generated functional logic program is analyzed
with the tool described in \citep{Hanus26SCICO}
to check which operations are fail-free.\footnote{The inclusion
of automated termination checks is currently omitted
since it is seldom that operations generated from Prolog
are completely defined but non-terminating.}
\item
Using the information computed in the previous step,
the given logic program is translated into a Curry program
based on the fail-sensitive functional transformation.
Thus, the same program structure is generated but the standard
function application $f\;e$ is replaced by \code{$f\;\$$!$\;e$}
whenever the expressions $e$ might fail, i.e.,
contains an operation which is not fail-free.
Note that $f$ might be an expression, e.g., another application.
For instance, if $f$ is a binary operation and $e_1$ as well as $e_2$
contain operations with non-trivial call types,
then an application $f~e_1~e_2$ generated in step 1 is replaced by
\code{($f\;\$$!$\;e_1$)$\;\$$!$\;e_2$}.\footnote{%
Note that this might change the order of evaluation
if both arguments are demanded by $f$,
but this order is not relevant in a declarative language without
side effects.}
If all operations in $e_2$ have trivial call types,
the application \code{($f\;\$$!$\;e_1$)$\;e_2$} is generated.
\end{enumerate}
A web interface to this transformation system is available at
\begin{center}
\url{https://cpm.curry-lang.org/webapps/pl2curry/}
\end{center}
This web page also contains all examples discussed in this paper.

\subsection{Extensions}
\label{sec:extensions}

Although our transformation tool is intended to transform logic programs,
it supports also a few extensions of pure logic programming
in order to experiment with more programs.

For instance, Prolog's built-in arithmetic predicates are useful
to avoid computing with numbers in Peano arithmetic.
For instance, the predicate \ccode{is} evaluates its second argument
to a number and returns this result by unifying it with its
first argument. If we use the result argument positions
$\respos(\code{is}/2) = \{1\}$,
the demand functional transformation replaces the literal
\code{M\;is\;N+1} by the local binding \code{M\;=\;N+1}.

Arbitrary occurrences of Prolog's ``cut'' operator
cannot be translated (or only into awkward code).
However, Prolog's if-then-else construct, often used as a restricted
form of a cut, can be translated in some situations.
If the condition is a simple predicate, like
a deterministic test or an arithmetic comparison,
it can be translated into a functional if-then-else construct.
For instance, consider the following definition of the
factorial function as a Prolog predicate with Prolog's built-in arithmetic:
\begin{prolog}
fac(N,F) :- (N=0 -> F=1 ; N1 is N - 1, fac(N1, F1), F is F1 * N).
\end{prolog}
Translating this with the demand functional transformation
and inlining local declarations,
we obtain the functional definition
\begin{curry}
fac n = if n == 0 then 1 else fac (n - 1) * n
\end{curry}
These extensions support the transformation
of many other predicates defining arithmetic functions.
Nevertheless, our tool can not translate arbitrary Prolog programs.
This is due to the fact that
Prolog programs often use impure features which cannot be
directly mapped into a language like Curry.
This is intended since functional logic languages avoid
impure features and side effects by concepts from functional programming.
For instance, I/O operations, offered in Prolog as
predicates with side effects, can be represented in
functional (logic) languages by monadic operations
which structure effectful computations \citep{Wadler97}.
Moreover, encapsulated search (\code{findall}) or cuts in Prolog,
whose behavior depends on the search strategy and ordering of rules,
can be represented in functional logic programming
in a strategy-independent manner as
set functions \citep{AntoyHanus09} or
default rules \citep{AntoyHanus17TPLP}.
Since the type system of Curry ensures that these extensions
are used in a declarative manner, an automatic transformation
of Prolog programs seems impossible since Prolog does not enforce
any restriction on the use of non-declarative features.

Although our transformation tool is not intended as a translator
for arbitrary Prolog programs, we evaluate our approach
by applying our tool to some Prolog programs
which do not use non-declarative features.

\section{Evaluation}
\label{sec:eval}

The main motivation for this work is to show that
functional logic programs have concrete operational advantages
compared to pure logic programs.
In principle, this has already been shown in \citep{Hanus22TPLP}
but that approach is based on changing the semantics of programs
from strict to demand-driven evaluation.
As discussed above, this could have the effect that
the transformed program computes \emph{more} answers
than the original logic program.
By incorporating failure information,
our new transformation does not have this effect but
computes only more general answers
compared to the answers of the logic program.
For instance, the Prolog goal
\begin{prolog}
?- plus(X,o,N), signat(N,s(o)).
\end{prolog}
yields an infinite set of answers where \code{X} and \code{N}
are bound to all non-zero natural numbers, whereas
the equivalent Curry expression obtained by our transformation
\begin{curry}
> S O =:= signat (plus x O)
\end{curry}
has a finite search space and yields the single answer substitution
$\{ \code{x} \mapsto \code{S$\;$\us}\}$.

In the following, we want to show the practical advantages
of this transformation by various examples.
Note that these examples are small programs.
Larger Prolog programs are seldom logic programs---they often
use non-logical features.
Such non-declarative features are either not necessary
in functional logic programs (e.g., cuts are replaced
by exploiting functional dependencies) or can be reformulated
in a declarative manner (e.g., declarative monadic I/O, state monads).
Due to these reasons, functional logic programs have
advantages from the view of software construction---which is
less tangible to formalization.
Therefore, we evaluate the advantages of functional logic programming
only from the operational point of view.
Since we want to demonstrate the operational improvements
of \emph{logic} programs due to the use of functions,
it is useless to consider non-logical features in this comparison.

\begin{table}[t]
\centering
\caption{Execution and transformation times of Prolog and generated Curry programs}
\label{table:benchmarks}
\begin{tabular}{l@{~~~}rrrc@{}c}
\topline
Language: & \multicolumn{1}{c}{Prolog}  & \multicolumn{1}{c}{Prolog}
          & \multicolumn{1}{c}{Curry} & Transform & Analysis \\
System:   & SWI 9.0.4~~ & SICStus 4.10.0~~ & KiCS2 3.5.0 & time & time
\midline
\code{rev{\char95}4096}                        &    0.19~~~~ &    0.18~~~~~ & 0.08~~~ & 1 & 27 \\
\code{takPeano{\char95}27{\char95}16{\char95}8}&    5.85~~~~ &    2.81~~~~~ & 0.57~~~ & 1 & 44 \\
\code{takInt{\char95}27{\char95}16{\char95}8}  &    0.47~~~~ &    0.21~~~~~ & 0.39~~~ & 1 & 28 \\
\code{ackermann{\char95}3{\char95}9}           &    1.93~~~~ &    9.68~~~~~ & 0.05~~~ & 1 & 30 \\
\code{primes{\char95}50000}                    &    6.23~~~~ &    6.50~~~~~ & 2.48~~~ & 1 & 42 \\
\code{qsort{\char95}6400}                      &    0.38~~~~ &    0.44~~~~~ & 0.07~~~ & 1 & 29 \\
\code{thirdof{\char95}0}                       &$\infty$~~~~ &$\infty$~~~~~ & 0.01~~~ & 1 & 38 \\
\code{signat{\char95}plus{\char95}0}           &$\infty$~~~~ &$\infty$~~~~~ & 0.01~~~ & 1 & 37 \\
\code{numnodes{\char95}7}                      &$\infty$~~~~ &$\infty$~~~~~ & 0.01~~~ & 1 & 40 \\
\code{sublist{\char95}12}                      &$\infty$~~~~ &$\infty$~~~~~ & 0.01~~~ & 1 & 38 \\
\code{pali{\char95}12321}                      &$\infty$~~~~ &$\infty$~~~~~ & 0.01~~~ & 1 & 38 \\
\code{gen{\char95}exp{\char95}5}               &$\infty$~~~~ &$\infty$~~~~~ & 0.06~~~ & 1 & 41 \\
\code{permsort{\char95}10}                     &    1.13~~~~ &    0.26~~~~~ & 0.02~~~ & 1 & 39 \\
\code{permsort{\char95}11}                     &   12.72~~~~ &    1.36~~~~~ & 0.06~~~ & 1 & 39 \\
\code{permsort{\char95}12}                     &  164.13~~~~ &   15.22~~~~~ & 0.22~~~ & 1 & 39 \\
\code{queens{\char95}8}                        &    0.02~~~~ &    0.16~~~~~ & 0.40~~~ & 2 & 42 \\
\code{send{\char95}more{\char95}money}         &    0.01~~~~ &    0.15~~~~~ & 0.54~~~ & 1 & 57 \\
\code{zebra}                                   &    0.01~~~~ &    0.15~~~~~ & 0.14~~~ & 1 & 147
\botline
\end{tabular}
\end{table}

Table~\ref{table:benchmarks} contains the results of executing
various Prolog programs with SWI-Prolog and SICStus-Prolog
and the Curry programs obtained by applying the
fail-sensitive functional transformation with
the Curry system KiCS2 \citep{BrasselHanusPeemoellerReck11}.
All benchmarks can be found in the tool implementation
(see Section~\ref{sec:toolimpl}).
The code of most benchmarks is also shown in \ref{app:benchmarks}.
The benchmarks were executed on a
Linux machine 
running Ubuntu 24.04 with an Intel Core Ultra 7 255U (4.2GHz) processor
with 12 cores and 32 GB memory.
The execution times are the total run time (in seconds) of executing the
binary generated with the respective Prolog and Curry systems.
The last two columns contain the times (in milliseconds) to transform
the Prolog programs into a Curry program and to analyze the generated
Curry program with the tool described in \citep{Hanus26SCICO}
to infer which operations are fail-free
(the tools were compiled with the Curry system KiCS2).
These columns show that the time to transform the Prolog programs
is negligible (around one millisecond) whereas the analysis times
are a bit higher (due to potential fixpoint computations)
but still acceptable.

The direct comparison of original and transformed programs
is not straightforward since the execution times depend
on the implementation techniques used in the Prolog and Curry systems.
KiCS2 compiles Curry programs into Haskell programs
that are compiled to machine code by the Glasgow Haskell Compiler
(GHC 9.4.5). GHC generates efficient code for functional computations.
This is visible in the first six benchmarks which
are purely deterministic computations.
\code{rev\us{}4096} is the naive list reversal applied to a
list of 4096 elements.
\code{takPeano\us{}27\us{}16\us{}8} applies
the highly recursive \code{tak} function, used in various
benchmarks \citep{Partain92} for logic and functional languages,
to the values \code{(27,16,8)} in Peano representation,
and \code{takInt\us{}27\us{}16\us{}8} computes the same
but with built-in integer arithmetic.
The Ackermann function is defined on Peano numbers
and applied to the Peano representation of \code{(3,9)}.
\code{primes{\char95}50000} computes the number of prime
numbers which are smaller 50000 via the sieve of Eratosthenes.
\code{qsort{\char95}6400} is the implementation of quicksort
based on partitioning a list into smaller and larger elements,
as defined by \citet{SterlingShapiro94}, applied to a list
of 6400 elements.

It is interesting to note that the definition of \code{takPeano}
decrements some arguments based on the predicate \code{dec}
defined in Section~\ref{ex:dec}.
Since the generated function \code{dec} might fail,
our transformation inserts the strict application
operator \ccode{$\$$!} in several places in the body of \code{takPeano}.
These operators cause some overhead\footnote{The operator
\ccode{$\$$!} is not considered by the Curry compiler KiCS2
but implemented by a specific predefined operation.}
and are not really necessary since
\code{takPeano} is strict in its first argument.
With a strictness analysis, one could drop these operators.
Although this is not implemented in our tool,
a manual optimization of the Curry code results in an
execution time of 0.36 seconds.
Thus, there is potential to improve the generated Curry code.

For these benchmarks, the demand-driven evaluation
strategy has no real advantage since the values of all
subexpressions are required.
Hence, the same steps, possibly in a different order, are performed
in the Prolog and Curry programs.
The situation is different in the next six benchmarks.
The first three of them are examples already discussed in this paper.
\code{thirdof\us{}O} is the evaluation of the literal
\code{thirdof(o,T)} (see Section~\ref{ex:thirdof}),
\code{signat\us{}plus\us{}O} is the evaluation of the goal
shown at the beginning of this section,
and \code{numnodes\us{}7} is the evaluation of the literal
\code{numnodes(T,s(s(s(s(s(s(s(o))))))))} having five solutions
(compare Example~\ref{ex:numnodes}).
Furthermore, \code{sublist{\char95}12} computes all sublists
of the list \code{[1,2]} by splitting the argument with
an inverse use of concatenating three lists,
\code{pali{\char95}12321} checks whether the
argument list \code{[1,2,3,2,1]} is a palindrome and relates
it to the middle element (defined by splitting
the argument list into three sublists and checking whether
the first is the reversal of the third),
and \code{gen{\char95}exp{\char95}5} generates arithmetic expressions
with five tokens based on a simple syntax description,
as defined by \citet{GonzalezEtAl99}.
Since Table~\ref{table:benchmarks} shows the time to compute
\emph{all} answers to the given goals,
the Prolog systems would not terminate due to the infinite
search spaces of these goals. Thus, these examples were not executed
but marked by ``$\infty$'' in the table.
In contrast, the transformed Curry programs have finite search spaces.

The \code{permsort\us$n$} benchmarks, where permutation sort is applied
to lists containing 10, 11, and 12 decreasing Peano numbers,
demonstrate the advantage of demand-driven
evaluation even if the search space is finite.
As discussed at various places \citep{AntoyHanus10CACM,GonzalezEtAl99,Hanus13},
the functional logic version of permutation sort has the effect
that permutations are explored in a demand-driven manner
so that not all permutations are actually generated.
Thus, our transformation maps a ``generate-and-test'' algorithm
into a more efficient ``test-of-generate-as-demanded'' algorithm
with a lower complexity, as apparent from the benchmarks.

The benchmark \code{queens\us{}8} computes solutions to place
eight queens on a chess board by incrementally adding one queen
at a time, as shown by \citet{SterlingShapiro94},
\code{send{\char95}more{\char95}money} solves
a crypto-arithmetic puzzle by guessing digits for letters,
and \code{zebra} solves a logic puzzle presented by \citet{SterlingShapiro94}.
These benchmarks show the influence of the implementation techniques
used in a concrete system to the absolute run times.
In contrast to almost all other benchmarks, SWI-Prolog is
the fastest whereas KiCS2 is the slowest.
This can be explained by the fact that
SWI-Prolog has a native implementation of backtracking
search whereas KiCS2 implements non-determinism via
the explicit construction of search trees to support
various search strategies by different search tree traversals.
The explicit construction of search trees causes some overhead
for programs where non-deterministic search dominates.

In general, the fail-sensitive functional transformation produces
functional logic programs providing the same or more general answers
as the original logic programs.
In the worst case, the same number of evaluation steps
are performed (apart from a few additional steps to reduce
occurrences of \ccode{$\$$!} introduced due to non-total functions).
In the best case, the transformation reduces infinite search spaces
to finite ones.
Thus, one does not get any disadvantage of this transformation
but in some cases considerable advantages.

We want to remark that this improvement is not only of
theoretical interest.
As apparent from our examples, logic programs might
have infinite search spaces when search is nested,
i.e., if two predicates defined on infinite structures,
like Peano numbers, lists, or trees, are sequentially
called with unknown arguments, as in \code{plus3} or \code{numnodes}.
When such programs are transformed into functional logic programs,
a demand-driven evaluation, like needed narrowing,
results in a demand-driven exploration of the search space.
This is exploited in \citep{Hanus11ICLP} to implement
a domain-specific language for deep XML matching and transformation
as a library in Curry.
Without the demand-driven evaluation strategy,
many of the library functions would not terminate.
Actually, the library has similar features as the
logic-based language Xcerpt \citep{BrySchaffert02ICLP}
which uses a specialized unification procedure to ensure
finite matching and unification of XML terms.

\begin{table}[t]
\centering
\caption{Memory usage of Prolog and Curry programs}
\label{table:benchmarks-memory}
\begin{tabular}{l@{~~~}rr@{\qquad}rr@{\qquad}rr}
\topline
Language: & \multicolumn{2}{c}{Prolog~~~~~}  & \multicolumn{2}{c}{Prolog}
          & \multicolumn{2}{c}{Curry} \\
System:   & \multicolumn{2}{c}{SWI 9.0.4~~~~~} & \multicolumn{2}{c}{SICStus 4.10.0~~}
          & \multicolumn{2}{c}{KiCS2 3.5.0}
\midline
\code{rev{\char95}4096}                        & 224 &   11900 & 44 &  20572 & 28558 & 16152 \\
\code{takPeano{\char95}27{\char95}16{\char95}8}& 224 & 1973532 & 45 & 727332 & 28555 & 12576 \\
\code{takInt{\char95}27{\char95}16{\char95}8}  & 223 &    8944 & 44 &  20612 & 28512 & 13672 \\
\code{ackermann{\char95}3{\char95}9}           & 223 & 1619020 & 44 & 300324 & 28557 & 19440 \\
\code{primes{\char95}50000}                    & 223 & 4409448 & 44 & 769424 & 28582 & 35152 \\
\code{qsort{\char95}6400}                      & 224 &  283812 & 44 &  80780 & 28598 & 32304 \\
\code{thirdof{\char95}0}                       & 223 &    n/a~ & 44 &   n/a~ & 28545 &  7660 \\
\code{signat{\char95}plus{\char95}0}           & 223 &    n/a~ & 44 &   n/a~ & 28545 &  7776 \\
\code{numnodes{\char95}7}                      & 223 &    n/a~ & 44 &   n/a~ & 28672 &  8752 \\
\code{sublist{\char95}12}                      & 223 &    n/a~ & 44 &   n/a~ & 28487 &  9040 \\
\code{pali{\char95}12321}                      & 223 &    n/a~ & 44 &   n/a~ & 28495 &  8884 \\
\code{gen{\char95}exp{\char95}5}               & 223 &    n/a~ & 44 &   n/a~ & 28583 & 26404 \\
\code{permsort{\char95}10}                     & 223 &    8772 & 44 &  20648 & 28574 & 19044 \\
\code{permsort{\char95}11}                     & 223 &    8904 & 44 &  20672 & 28574 & 31340 \\
\code{permsort{\char95}12}                     & 223 &    8944 & 44 &  20716 & 28574 & 70272 \\
\code{queens{\char95}8}                        & 224 &    9072 & 45 &  20608 & 28586 & 17116 \\
\code{send{\char95}more{\char95}money}         & 224 &    8824 & 44 &  20664 & 28600 & 31580 \\
\code{zebra}                                   & 224 &    8812 & 44 &  20548 & 28875 & 15228
\botline
\end{tabular}
\end{table}

Finally, we evaluate the memory footprint of the benchmark programs.
Since they are executed with Prolog and Curry systems
having different implementation techniques (e.g., different run-time
structures, different memory management and garbage collection),
their actual memory footprints are not easy to compare.
Nevertheless, Table~\ref{table:benchmarks-memory} shows some numbers
about the memory usage of the benchmark programs.

For each system, the first column shows the size (in kilobytes)
of the executable generated by the system.
The second column shows the maximum resident set size of the
process executing this process (in kilobytes) measured with
the GNU/Linux \code{time} command.
Since this maximum cannot be determined for
benchmarks with a conceptually infinite execution time,
the table contains ``n/a'' at these positions.

The comparison of the first columns shows that the executables
generated by KiCS2 are much larger than the executables
generated by the Prolog systems.
This is mainly due to the fact that the Curry executables
contain the translated code of the standard prelude of Curry
which comprises arithmetic and I/O operations as well as
many predefined list operations. Altogether, the prelude contains
more than one thousand (public and internal) operations.
Since the KiCS2 compiler translates each Curry operation
into an enhanced Haskell operation which is compiled into machine
code by the Haskell compiler, the generated code is much larger
compared to Prolog.
Fortunately, as shown in Table~\ref{table:benchmarks}, the size of the
executables is not really relevant to the execution times.

The comparison of the second columns does not allow a clear interpretation.
In some cases, the memory footprint of SWI-Prolog is small,
whereas in other cases, the memory footprints of SICStus-Prolog
or KiCS2 are smaller.

\section{Related Work}
\label{sec:related}

The view that algorithms can be seen
as declarative logic descriptions combined with appropriate
control rules was introduced decades ago \citep{Kowalski79}.
Since the standard control rule of Prolog, left-to-right
depth-first search implemented via backtracking,
has the risk to loop infinitely instead of delivering
an answer,
there is a long history to modify the standard control rule
in order to improve the operational behavior
\citep{BruynoogheEtAl89,Narain86}.
These proposals usually consider the operational level
so that a declarative justification
(soundness and completeness) is missing.
This is in contrast to our work which is motivated
to keep the soundness and completeness of logic programming
and provide a better operational model.

There are also approaches to add functions and functional notations
to logic programs and map them in various ways to logic programs,
in particular, to Prolog systems with coroutining in order
to have a similar effect as in a demand-driven computation model.
However, one has to be careful since there are various ways
and pitfalls to implement laziness in this way.
For instance, \citet{Naish91} implements lazy evaluation in Prolog
by representing closures as terms and use \code{when} declarations
to delay insufficiently instantiated function calls.
This might lead to floundering so that
completeness is lost when predicates are transformed into functions.
Moreover, delaying recursively defined predicates could result
in infinite search spaces where a complete narrowing strategy
has a finite search space \citep{Hanus95JLP}.
A notation for functions in Prolog programs is also proposed
in \citep{CasasCabezaHermenegildo06}. This syntactic extension
is mapped into Prolog predicates where coroutining is used to implement
lazy evaluation.
Although this syntactic transformation might yield the same values
and search space as functional logic languages, there are no formal results
justifying this transformation.

Other approaches use Prolog as a target language to implement
functional logic languages based on demand-driven narrowing strategies
\citep{AntoyHanus00FROCOS,Jimenez-MartinMarino-CarballoMoreno-Navarro92,%
LoogenLopezFraguasRodriguezArtalejo93PLILP,Lopez-FraguasSanchez-Hernandez99}.
Since these approaches implement narrowing strategies,
the soundness and completeness results of narrowing
holds also for the generated logic programs.
When implementing needed narrowing in this way \citep{AntoyHanus00FROCOS},
the resulting programs are optimal,
i.e., the set of computed solutions is minimal
and successful derivations have the shortest possible length
\citep{AntoyEchahedHanus00JACM}.

In this paper we take an opposite approach since we use
(pure) Prolog as the source language.
This was already proposed in \citep{Hanus22TPLP}
but there it is not ensured that the set of answers is kept
by the transformation, as discussed in Section~\ref{sec:lp2flp}.
By respecting the failure behavior of the source programs,
we fixed this problem and showed how to obtain an
equal or better operational behavior without changing the
answer behavior.

\section{Conclusions}
\label{sec:concl}

We presented a method to transform logic programs
into functional logic programs
so that the transformed functional logic program
always computes the same or more general answers
than the original program.
For a well-defined class of programs
(predicates completely defined by inductively sequential rules),
the set of answers and the number of evaluation steps
leading to an answer are minimal
\citep{Antoy97ALP,AntoyEchahedHanus00JACM}.
Thus, the transformation is useful to transfer
strong results from functional logic programming into logic programming.

To ensure the correctness of the transformation,
we use an approximation of the failure behavior of transformed programs,
which is improved by considering types provided for logic programs.
As a result, the transformation yields programs that
require the same number of evaluation steps in the worst case
but often require less steps and reduces infinite search spaces
to finite ones.
Thus, the main lesson learned from this work is that
the usage of functions, even in a logic programming manner
as supported in functional logic programming,
has many advantages but no disadvantage compared to
pure logic programming based solely on relations.
This is due to the fact that functional dependencies
between arguments and result values can be exploited
to provide a more efficient operational behavior
thanks to the demand-driven evaluation of nested
function calls.
We implemented our method in a tool and showed the practical
advantages by evaluating a set of small but typical examples.
Note that these advantages do not hold for all functional logic languages.
For instance, the functional logic language
Verse \citep{AugustssonEtAl23} fixes a deterministic,
backtracking-like search strategy without a demand-driven evaluation
of expressions so that the results shown in this paper
do not transfer to Verse.

For future work it might be interesting to improve
the presented fail-sensitive functional transformation
by taking strictness information into account to avoid
the introduction of strict application operations,
as discussed in Section~\ref{sec:eval}.
However, the motivation of this work is not to provide
a general applicable tool to transform Prolog programs,
since this is difficult due to the use of non-declarative features
in larger Prolog programs.
Our intention is to show that the execution mechanism
of functional logic programs is always equal or better than
the resolution principle for pure logic programs
thanks to exploiting functional dependencies
for demand-driven evaluations.
Although this could be simulated in Prolog systems
with coroutining, additional problems might occur
due to floundering and incompleteness of executions.

Applications implemented with logic programming techniques
are usually not pure logic programs since they require
an interaction with the outside world, e.g., file systems,
graphical user interfaces, or networks.
For this purpose, Prolog offers predicates with side effects.
Since the influence of side effects to logic computations
is difficult to analyze, our transformation approach
covers only pure logic programs.
Non-declarative aspects of Prolog programs can be replaced
by declarative functional programming features,
like monadic I/O \citep{Wadler97}.
Since the structure of monadic programs is quite different
from Prolog programs with non-declarative predicates,
there does not exist a general method to transform real Prolog programs
into Curry programs.
In applications written in Curry, non-deterministic logic computations
are often separated from deterministic monadic computations
by encapsulating logic computations, e.g.,
\citep{AntoyBrassel07,AntoyHanus09}.
Our work is intended to show that these logic computations
can be better implemented in a functional logic manner.
Actually, this is exploited in various applications written in Curry,
like the curricula and module information system of our department%
\footnote{\scriptsize\url{https://moduldb.informatik.uni-kiel.de/}},
where XML documents are processed with a functional logic library
exploiting demand-driven search \citep{Hanus11ICLP},
or the Curry compiler RICE \citep{Libby23}
which exploits functional logic program transformations \citep{HanusLibby25WLP}
for code optimizations in order to produce efficient target code.

\paragraph{Acknowledgments.}
The author is grateful to the anonymous reviewers
for their suggestions to improve a previous version of this paper.

%


\begin{thebibliography}{}

\bibitem[A{\"\i}t-Kaci, 1991]{Ait-Kaci91}
{\sc A{\"\i}t-Kaci, H.} 1991.
\newblock {\em Warren's Abstract Machine}.
\newblock MIT Press.

\bibitem[Antoy, 1992]{Antoy92ALP}
{\sc Antoy, S.}
\newblock Definitional trees.
\newblock In {\em Proc. of the 3rd International Conference on Algebraic and
  Logic Programming} 1992, pp. 143--157. Springer LNCS 632.

\bibitem[Antoy, 1997]{Antoy97ALP}
{\sc Antoy, S.}
\newblock Optimal non-deterministic functional logic computations.
\newblock In {\em Proc. International Conference on Algebraic and Logic
  Programming (ALP'97)} 1997, pp. 16--30. Springer LNCS 1298.

\bibitem[Antoy and Bra{\ss}el, 2007]{AntoyBrassel07}
{\sc Antoy, S.} {\sc and} {\sc Bra{\ss}el, B.}
\newblock Computing with subspaces.
\newblock In {\em Proceedings of the 9th ACM SIGPLAN International Conference
  on Principles and Practice of Declarative Programming (PPDP'07)} 2007, pp.
  121--130. ACM Press.

\bibitem[Antoy et~al., 2000]{AntoyEchahedHanus00JACM}
{\sc Antoy, S.}, {\sc Echahed, R.}, {\sc and} {\sc Hanus, M.} 2000.
\newblock A needed narrowing strategy.
\newblock {\em Journal of the ACM}, {\it 47}, 4, 776--822.

\bibitem[Antoy and Hanus, 2000]{AntoyHanus00FROCOS}
{\sc Antoy, S.} {\sc and} {\sc Hanus, M.}
\newblock Compiling multi-paradigm declarative programs into {Prolog}.
\newblock In {\em Proc. International Workshop on Frontiers of Combining
  Systems (FroCoS'2000)} 2000, pp. 171--185. Springer LNCS 1794.

\bibitem[Antoy and Hanus, 2009]{AntoyHanus09}
{\sc Antoy, S.} {\sc and} {\sc Hanus, M.}
\newblock Set functions for functional logic programming.
\newblock In {\em Proceedings of the 11th ACM SIGPLAN International Conference
  on Principles and Practice of Declarative Programming (PPDP'09)} 2009, pp.
  73--82. ACM Press.

\bibitem[Antoy and Hanus, 2010]{AntoyHanus10CACM}
{\sc Antoy, S.} {\sc and} {\sc Hanus, M.} 2010.
\newblock Functional logic programming.
\newblock {\em Communications of the ACM}, {\it 53}, 4, 74--85.

\bibitem[Antoy and Hanus, 2017]{AntoyHanus17TPLP}
{\sc Antoy, S.} {\sc and} {\sc Hanus, M.} 2017.
\newblock Default rules for {Curry}.
\newblock {\em Theory and Practice of Logic Programming}, {\it 17}, 2,
  121--147.

\bibitem[Augustsson et~al., 2023]{AugustssonEtAl23}
{\sc Augustsson, L.}, {\sc Breitner, J.}, {\sc Claessen, K.}, {\sc Jhala, R.},
  {\sc Peyton~Jones, S.}, {\sc Shivers, O.}, {\sc Steele, G.}, {\sc and} {\sc
  Sweeney, T.}
\newblock The {Verse} calculus: A core calculus for deterministic functional
  logic programming.
\newblock In {\em Proc.\ ACM International Conference on Functional Programming
  (ICFP 2023)} 2023, pp. 203:1--203:31.

\bibitem[Baader and Nipkow, 1998]{BaaderNipkow98}
{\sc Baader, F.} {\sc and} {\sc Nipkow, T.} 1998.
\newblock {\em Term Rewriting and All That}.
\newblock Cambridge University Press.

\bibitem[Barbosa et~al., 2021]{BarbosaFloridoSantosCosta21}
{\sc Barbosa, J.}, {\sc Florido, M.}, {\sc and} {\sc Santos~Costa, V.}
\newblock Data type inference for logic programming.
\newblock In {\em Proceedings of the 31st International Symposium on
  Logic-based Program Synthesis and Transformation (LOPSTR 2021)} 2021, pp.
  16--37. Springer LNCS 13290.

\bibitem[Barbuti et~al., 1984]{BarbutiBelliaLeviMartelli84}
{\sc Barbuti, R.}, {\sc Bellia, M.}, {\sc Levi, G.}, {\sc and} {\sc Martelli,
  M.}
\newblock On the integration of logic programming and functional programming.
\newblock In {\em Proc. IEEE Internat. Symposium on Logic Programming} 1984,
  pp. 160--166, Atlantic City.

\bibitem[Bra{\ss}el et~al., 2011]{BrasselHanusPeemoellerReck11}
{\sc Bra{\ss}el, B.}, {\sc Hanus, M.}, {\sc Peem{\"o}ller, B.}, {\sc and} {\sc
  Reck, F.}
\newblock {KiCS2}: A new compiler from {Curry} to {Haskell}.
\newblock In {\em Proc. of the 20th International Workshop on Functional and
  (Constraint) Logic Programming (WFLP 2011)} 2011, pp. 1--18. Springer LNCS
  6816.

\bibitem[Bruynooghe et~al., 1989]{BruynoogheEtAl89}
{\sc Bruynooghe, M.}, {\sc De~Schreye, D.}, {\sc and} {\sc Krekels, B.} 1989.
\newblock Compiling control.
\newblock {\em Journal of Logic Programming}, {\it 6}, 1-2, 135--162.

\bibitem[Bry and Schaffert, 2002]{BrySchaffert02ICLP}
{\sc Bry, F.} {\sc and} {\sc Schaffert, S.}
\newblock Towards a declarative query and transformation language for {XML} and
  semistructured data: Simulation unification.
\newblock In {\em Proceedings of the International Conference on Logic
  Programming (ICLP'02)} 2002, pp. 255--270. Springer LNCS 2401.

\bibitem[Casas et~al., 2006]{CasasCabezaHermenegildo06}
{\sc Casas, A.}, {\sc Cabeza, D.}, {\sc and} {\sc Hermenegildo, M.}
\newblock A syntactic approach to combining functional notation, lazy
  evaluation, and higher-order in {LP} systems.
\newblock In {\em Proc.\ of the 8th International Symposium on Functional and
  Logic Programming (FLOPS 2006)} 2006, pp. 146--162. Springer LNCS 3945.

\bibitem[Giesl et~al., 2011]{GieslEtAl11TOPLAS}
{\sc Giesl, J.}, {\sc Raffelsieper, M.}, {\sc Schneider-Kamp, P.}, {\sc
  Swiderski, S.}, {\sc and} {\sc Thiemann, R.} 2011.
\newblock Automatic termination proofs for {Haskell} by term rewriting.
\newblock {\em ACM Transactions on Programming Languages and Systems}, {\it
  33}, 2, Article 7.

\bibitem[Giovannetti et~al., 1991]{GiovannettiLeviMoisoPalamidessi91}
{\sc Giovannetti, E.}, {\sc Levi, G.}, {\sc Moiso, C.}, {\sc and} {\sc
  Palamidessi, C.} 1991.
\newblock Kernel leaf: A logic plus functional language.
\newblock {\em Journal of Computer and System Sciences}, {\it 42}, 2, 139--185.

\bibitem[Gonz{\'a}lez-Moreno et~al., 1999]{GonzalezEtAl99}
{\sc Gonz{\'a}lez-Moreno, J.}, {\sc Hortal{\'a}-Gonz{\'a}lez, M.}, {\sc
  L{\'o}pez-Fraguas, F.}, {\sc and} {\sc Rodr{\'\i}guez-Artalejo, M.} 1999.
\newblock An approach to declarative programming based on a rewriting logic.
\newblock {\em Journal of Logic Programming}, {\it 40}, 47--87.

\bibitem[Hanus, 1994]{Hanus94JLP}
{\sc Hanus, M.} 1994.
\newblock The integration of functions into logic programming: From theory to
  practice.
\newblock {\em Journal of Logic Programming}, {\it 19\&20}, 583--628.

\bibitem[Hanus, 1995]{Hanus95JLP}
{\sc Hanus, M.} 1995.
\newblock Analysis of residuating logic programs.
\newblock {\em Journal of Logic Programming}, {\it 24}, 3, 161--199.

\bibitem[Hanus, 2011]{Hanus11ICLP}
{\sc Hanus, M.}
\newblock Declarative processing of semistructured web data.
\newblock In {\em Technical Communications of the 27th International Conference
  on Logic Programming} 2011, volume~11, pp. 198--208. Leibniz International
  Proceedings in Informatics (LIPIcs).

\bibitem[Hanus, 2013]{Hanus13}
{\sc Hanus, M.}
\newblock Functional logic programming: From theory to {Curry}.
\newblock In {\em Programming Logics - Essays in Memory of Harald Ganzinger}
  2013, pp. 123--168. Springer LNCS 7797.

\bibitem[Hanus, 2018]{Hanus18PPDP}
{\sc Hanus, M.}
\newblock Verifying fail-free declarative programs.
\newblock In {\em Proceedings of the 20th International Symposium on Principles
  and Practice of Declarative Programming (PPDP 2018)} 2018, pp. 12:1--12:13.
  ACM Press.

\bibitem[Hanus, 2022]{Hanus22TPLP}
{\sc Hanus, M.} 2022.
\newblock From logic to functional logic programs.
\newblock {\em Theory and Practice of Logic Programming}, {\it 22}, 4,
  538--554.

\bibitem[Hanus, 2024]{Hanus24LOPSTR}
{\sc Hanus, M.}
\newblock Improving logic programs by adding functions.
\newblock In {\em Proceedings of the 34th International Symposium on
  Logic-Based Program Synthesis and Transformation (LOPSTR 2024)} 2024, pp.
  27--44. Springer LNCS 14919.

\bibitem[Hanus, 2026]{Hanus26SCICO}
{\sc Hanus, M.} 2026.
\newblock Inferring non-failure conditions for declarative programs.
\newblock {\em Science of Computer Programming}, {\it 250}, 103416.

\bibitem[Hanus and Libby, 2025]{HanusLibby25WLP}
{\sc Hanus, M.} {\sc and} {\sc Libby, S.}
\newblock Functional logic program transformations.
\newblock In {\em Proc.\ of the 39th Workshop on(Constraint) Logic Programming
  (WLP 2025)} 2025. To appear in Springer LNCS.

\bibitem[Hanus and Skrlac, 2014]{HanusSkrlac14}
{\sc Hanus, M.} {\sc and} {\sc Skrlac, F.}
\newblock A modular and generic analysis server system for functional logic
  programs.
\newblock In {\em Proc. of the ACM SIGPLAN 2014 Workshop on Partial Evaluation
  and Program Manipulation (PEPM'14)} 2014, pp. 181--188. ACM Press.

\bibitem[Hanus~{(ed.)}, 2016]{Hanus16Curry}
{\sc Hanus~{(ed.)}, M.} 2016.
\newblock Curry: An integrated functional logic language (vers.\ 0.9.0).
\newblock Available at \url{https://www.curry-lang.org}.

\bibitem[Hermenegildo et~al., 2012]{HermenegildoEtAl12}
{\sc Hermenegildo, M.}, {\sc Bueno, F.}, {\sc Carro, M.}, {\sc
  L{\'o}pez-Garc{\'\i}a, P.}, {\sc Mera, E.}, {\sc Morales, J.}, {\sc and} {\sc
  Puebla, G.} 2012.
\newblock An overview of {Ciao} and its design philosophy.
\newblock {\em Theory and Practice of Logic Programming}, {\it 12}, 1-2,
  219--252.

\bibitem[Huet and L{\'e}vy, 1991]{HuetLevy91}
{\sc Huet, G.} {\sc and} {\sc L{\'e}vy, J.-J.}
\newblock Computations in orthogonal rewriting systems.
\newblock In {\sc Lassez, J.-L.} {\sc and} {\sc Plotkin, G.}, editors, {\em
  Computational Logic: Essays in Honor of Alan Robinson} 1991, pp. 395--443.
  MIT Press.

\bibitem[Hughes, 1990]{Hughes90}
{\sc Hughes, J.}
\newblock Why functional programming matters.
\newblock In {\sc Turner, D.}, editor, {\em Research Topics in Functional
  Programming} 1990, pp. 17--42. Addison Wesley.

\bibitem[Jim{\'e}nez-Martin et~al.,
  1992]{Jimenez-MartinMarino-CarballoMoreno-Navarro92}
{\sc Jim{\'e}nez-Martin, J.}, {\sc Marino-Carballo, J.}, {\sc and} {\sc
  Moreno-Navarro, J.}
\newblock Efficient compilation of lazy narrowing into {Prolog}.
\newblock In {\em Proc. Int. Workshop on Logic Program Synthesis and
  Transformation (LOPSTR'92)} 1992, pp. 253--270. Springer Workshops in
  Computing Series.

\bibitem[Kowalski, 1979]{Kowalski79}
{\sc Kowalski, R.} 1979.
\newblock Algorithm = logic + control.
\newblock {\em Communications of the ACM}, {\it 22}, 7, 424--436.

\bibitem[Launchbury, 1993]{Launchbury93}
{\sc Launchbury, J.}
\newblock A natural semantics for lazy evaluation.
\newblock In {\em Proc. 20th ACM Symposium on Principles of Programming
  Languages (POPL'93)} 1993, pp. 144--154. ACM Press.

\bibitem[Lee et~al., 2001]{LeeJonesBenAmram01}
{\sc Lee, C.}, {\sc Jones, N.}, {\sc and} {\sc Ben-Amram, A.}
\newblock The size-change principle for program termination.
\newblock In {\em ACM Symposium on Principles of Programming Languages
  (POPL'01)} 2001, pp. 81--92.

\bibitem[Libby, 2023]{Libby23}
{\sc Libby, S.}
\newblock {RICE}: An optimizing {Curry} compiler.
\newblock In {\em 25th International Symposium Practical Aspects of Declarative
  Languages (PADL 2023)} 2023, pp. 3--19. Springer LNCS 13880.

\bibitem[Lloyd, 1987]{Lloyd87}
{\sc Lloyd, J.} 1987.
\newblock {\em Foundations of Logic Programming}.
\newblock Springer, second, extended edition.

\bibitem[Loogen et~al., 1993]{LoogenLopezFraguasRodriguezArtalejo93PLILP}
{\sc Loogen, R.}, {\sc L{\'o}pez~Fraguas, F.}, {\sc and} {\sc
  Rodr{\'\i}guez~Artalejo, M.}
\newblock A demand driven computation strategy for lazy narrowing.
\newblock In {\em Proc. of the 5th International Symposium on Programming
  Language Implementation and Logic Programming} 1993, pp. 184--200. Springer
  LNCS 714.

\bibitem[L\'opez-Fraguas and S\'anchez-Hern\'andez,
  1999]{Lopez-FraguasSanchez-Hernandez99}
{\sc L\'opez-Fraguas, F.} {\sc and} {\sc S\'anchez-Hern\'andez, J.}
\newblock {TOY}: A multiparadigm declarative system.
\newblock In {\em Proc. of RTA'99} 1999, pp. 244--247. Springer LNCS 1631.

\bibitem[Mycroft and O'Keefe, 1984]{MycroftOKeefe84}
{\sc Mycroft, A.} {\sc and} {\sc O'Keefe, R.} 1984.
\newblock A polymorphic type system for {Prolog}.
\newblock {\em Artificial Intelligence}, {\it 23}, 3, 295--307.

\bibitem[Naish, 1991]{Naish91}
{\sc Naish, L.}
\newblock Adding equations to {NU-Prolog}.
\newblock In {\em Proc. of the 3rd Int. Symposium on Programming Language
  Implementation and Logic Programming} 1991, pp. 15--26. Springer LNCS 528.

\bibitem[Narain, 1986]{Narain86}
{\sc Narain, S.} 1986.
\newblock A technique for doing lazy evaluation in logic.
\newblock {\em Journal of Logic Programming}, {\it 3}, 259--276.

\bibitem[Partain, 1992]{Partain92}
{\sc Partain, W.}
\newblock The nofib benchmark suite of {Haskell} programs.
\newblock In {\em Proceedings of the 1992 Glasgow Workshop on Functional
  Programming} 1992, pp. 195--202. Springer.

\bibitem[Peyton~Jones, 2003]{PeytonJones03Haskell}
{\sc Peyton~Jones, S.}, editor 2003.
\newblock {\em Haskell 98 Language and Libraries---The Revised Report}.
\newblock Cambridge University Press.

\bibitem[Pfenning, 1992]{Pfenning92}
{\sc Pfenning, F.}, editor 1992.
\newblock {\em Types in Logic Programming}.
\newblock MIT Press.

\bibitem[Reddy, 1984]{Reddy84}
{\sc Reddy, U.}
\newblock Transformation of logic programs into functional programs.
\newblock In {\em Proc. IEEE Internat. Symposium on Logic Programming} 1984,
  pp. 187--196, Atlantic City.

\bibitem[Reddy, 1985]{Reddy85}
{\sc Reddy, U.}
\newblock Narrowing as the operational semantics of functional languages.
\newblock In {\em Proc. IEEE Internat. Symposium on Logic Programming} 1985,
  pp. 138--151, Boston.

\bibitem[Sato and Tamaki, 1984]{SatoTamaki84}
{\sc Sato, T.} {\sc and} {\sc Tamaki, H.} 1984.
\newblock Enumeration of success patterns in logic programs.
\newblock {\em Theoretical Computer Science}, {\it 34}, 227--240.

\bibitem[Schrijvers et~al., 2008]{SchrijversSantosCostaWielemakerDemoen08}
{\sc Schrijvers, T.}, {\sc Santos~Costa, V.}, {\sc Wielemaker, J.}, {\sc and}
  {\sc Demoen, B.}
\newblock Towards {Typed Prolog}.
\newblock In {\em 24th International Conference on Logic Programming (ICLP
  2008)} 2008, pp. 693--697. Springer LNCS 5366.

\bibitem[Slagle, 1974]{Slagle74}
{\sc Slagle, J.} 1974.
\newblock Automated theorem-proving for theories with simplifiers,
  commutativity, and associativity.
\newblock {\em Journal of the ACM}, {\it 21}, 4, 622--642.

\bibitem[Sterling and Shapiro, 1994]{SterlingShapiro94}
{\sc Sterling, L.} {\sc and} {\sc Shapiro, E.} 1994.
\newblock {\em The Art of Prolog}.
\newblock MIT Press, Cambridge, Massachusetts, 2nd edition.

\bibitem[Van~Roy and Haridi, 2023]{VanRoyHaridi23}
{\sc Van~Roy, P.} {\sc and} {\sc Haridi, S.} 2023.
\newblock Ideas for the future of {Prolog} inspired by {Oz}.
\newblock {\em CoRR}, {\it abs/2302.00558}.

\bibitem[Wadler, 1997]{Wadler97}
{\sc Wadler, P.} 1997.
\newblock How to declare an imperative.
\newblock {\em ACM Computing Surveys}, {\it 29}, 3, 240--263.

\bibitem[Warren, 1983]{Warren83}
{\sc Warren, D.} 1983.
\newblock An abstract {Prolog} instruction set.
\newblock Technical note 309, SRI International, Stanford.

\end{thebibliography}

\newpage
\appendix

\section{Benchmark Programs}
\label{app:benchmarks}

This appendix shows the Prolog source code of some predicates
used in the benchmarks in Section~\ref{sec:eval}
and the Curry code generated by our tool
implementing the fail-sensitive functional transformation
(with inlining for better readability).

\subsection{Predicate \texttt{rev}}\label{app:rev}

The predicate \code{rev} is the well-known naive reverse
with a quadratic complexity:
\begin{prolog}
app([],Xs,Xs).
app([X|Xs],Ys,[X|Zs]) :- app(Xs,Ys,Zs).$\listline$
rev([],[]).
rev([X|Xs],R) :- rev(Xs,Zs), app(Zs,[X],R).
\end{prolog}
Both predicates are translated into totally defined functions:
\begin{curry}
app [] xs = xs
app (x : xs) ys = x : app xs ys$\listline$
rev [] = []
rev (x : xs) = app (rev xs) [x]
\end{curry}

\subsection{Predicate \texttt{tak} on Peano numbers}

The function \code{tak} is defined in Prolog on Peano numbers,
i.e., terms constructed from \code{o/0} and \code{s/1}:
\begin{prolog}
tak(X,Y,Z,A) :-	leq(X,Y,XLEQY), takc(XLEQY,X,Y,Z,A).

takc(true,X,Y,Z,Z).
takc(false,X,Y,Z,A) :-
        dec(X,X1),
        tak(X1,Y,Z,A1),
        dec(Y,Y1),
        tak(Y1,Z,X,A2),
        dec(Z,Z1),
        tak(Z1,X,Y,A3),
        tak(A1,A2,A3,A).

dec(s(X),X).

leq(o,_,true).
leq(s(_),o,false).
leq(s(X),s(Y),R) :- leq(X,Y,R).
\end{prolog}
Due to the definition of \code{dec}, the generated function
\code{takc} contains occurrences of \ccode{$\$$!} in its right-hand side:
\begin{curry}
tak x y z = takc (leq x y) x y z

takc True x y z = z
takc False x y z =
  ((tak $\$$! (tak $\$$! (dec x)) y z) $\$$! (tak $\$$! (dec y)) z x) $\$$!
     (tak $\$$! (dec z)) x y

dec (S x) = x

leq O _ = True
leq (S _) O = False
leq (S x) (S y) = leq x y
\end{curry}

\subsection{Predicate \texttt{tak} with built-in integer arithmetic}

This is similarly to the previous example but defined
with built-in integer arithmetic:
\begin{prolog}
tak(X,Y,Z,A) :-
	X =< Y
        -> A=Z
	 ; ((X1 is X - 1),
            tak(X1,Y,Z,A1),
            (Y1 is Y - 1),
            tak(Y1,Z,X,A2),
            (Z1 is Z - 1),
            tak(Z1,X,Y,A3),
            tak(A1,A2,A3,A)).
\end{prolog}
Due to extensions of the transformation tool to built-in integer
arithmetic and Prolog's if-then-else construct
(see Section~\ref{sec:extensions}),
this example is transformed into
\begin{curry}
tak x y z =
  if x <= y
     then z
     else tak (tak (x - 1) y z) (tak (y - 1) z x) (tak (z - 1) x y)
\end{curry}

\subsection{Predicate \texttt{ackermann}}

The Ackermann function is defined as a Prolog predicate on Peano numbers,
as presented by \citet{SterlingShapiro94}:
\begin{prolog}
ackermann(o,N,s(N)).
ackermann(s(M),o,Val) :- ackermann(M,s(o),Val).
ackermann(s(M),s(N),Val) :-
  ackermann(s(M),N,Val1), ackermann(M,Val1,Val).
\end{prolog}
It is translated into the Curry function
\begin{curry}
ackermann O n = S n
ackermann (S m) O = ackermann m (S O)
ackermann (S m) (S n) = ackermann m (ackermann (S m) n)
\end{curry}

\subsection{Predicates \texttt{sublist} and \texttt{pali}}\label{app:pali}

The predicate \code{sublist} computes all sublists of a list
by splitting the argument into three lists.
The predicate \code{pali} checks whether the first argument is a palindrome
and relates it to its middle element, i.e., the palindrome must have
an odd length. This is implemented by using the predicates
\code{app} and \code{rev} (see Section~\ref{app:rev})
and splitting the argument list into three sublists and checking whether
the first is the reversal of the third.
The splitting of a list into three sublists uses the predicate \code{app3}
which relates three lists to its concatenated lists.
\begin{prolog}
app3(Xs,Ys,Zs,Ts) :- app(Xs,Ys,Rs), app(Rs,Zs,Ts).

sublist(Xs,Sub) :- app3(_,Sub,_,Xs).

pali(Zs,X) :- app3(Xs,[X],Ys,Zs), rev(Xs,Ys).
\end{prolog}
These predicates are translated into the Curry functions
\begin{curry}
app3 xs ys zs = app (app xs ys) zs

sublist xs | xs =:= app3 _ sub _ = sub
  where sub free

pali zs | zs =:= app3 xs [x] (rev xs) = x
  where xs, x free
\end{curry}
\ccode{?-$\;$sublist([1,2],S).} has six answers
and the query \ccode{?-$\;$pali([1,2,3,2,1],M).}
has one answer \code{M=3}, but both causes infinite search spaces
in Prolog, whereas the expressions \ccode{sublist$\;$[1,2]}
and \ccode{pali$\;$[1,2,3,2,1]}
have finite search spaces.

\subsection{Predicate \texttt{permsort}}

The permutation sort example computes permutations
by non-deterministically inserting an element into a list.
\begin{prolog}
% Non-deterministic list insertion:
insert(X,[],[X]).
insert(X,[Y|Ys],[X,Y|Ys]).
insert(X,[Y|Ys],[Y|Zs]) :- insert(X,Ys,Zs).

% Permutations:
perm([],[]).
perm([X|Xs],Zs) :- perm(Xs,Ys), insert(X,Ys,Zs).

% less-or-equal relation
leq(o,_).
leq(s(X),s(Y)) :- leq(X,Y).

% Is the argument list sorted?
sorted([]).
sorted([_]).
sorted([X,Y|Ys]) :- leq(X,Y), sorted([Y|Ys]).

% Permutation sort: search for some sorted permutation
psort(Xs,Ys) :- perm(Xs,Ys), sorted(Ys).
\end{prolog}
The generated operations \code{insert} and \code{perm} are
totally defined, whereas \code{leq}, \code{sorted}, and \code{psort}
might fail.
Due to the strict left-to-right semantics of the predefined
conjunction operator \ccode{\&\&},
insertions of the strict application operator \ccode{$\$$!}
in the third rule of \code{sorted} are not necessary.
\begin{curry}
insert x [] = [x]
insert x (y : ys) = x : (y : ys)
insert x (y : ys) = y : insert x ys

perm [] = []
perm (x : xs) = insert x (perm xs)

leq O _ = True
leq (S x) (S y) | leq x y = True

sorted [] = True
sorted [_] = True
sorted (x : y : ys) | leq x y && sorted (y : ys) = True

psort xs | sorted ys = ys
  where ys = perm xs
\end{curry}

\end{document}